%% file: ms.tex
\documentclass[iop]{emulateapj}
\slugcomment{{\sc Accepted to ApJ:}August 21, 2012} 
\usepackage{graphics}
\pdfoutput=1

\newcommand{\ltsimeq}{\la}
\newcommand{\gtsimeq}{\ga}
\newcommand{\lsun}{L$_{\odot}$}
\newcommand{\msun}{M$_{\odot}$}

\newcommand{\HII}{H~{\sc ii}}

\shortauthors{McQuinn et al.}
\shorttitle{Spatial Distribution of Star Formation}

\begin{document}
\title{The Nature of Starbursts: III. The Spatial Distribution of Star Formation\footnote{Based on observations made with the NASA/ESA Hubble Space Telescope, obtained from the Data Archive at the Space Telescope Science Institute, which is operated by the Association of Universities for Research in Astronomy, Inc., under NASA contract NAS 5-26555.}}
\author{Kristen~B.~W. McQuinn\altaffilmark{1}, 
Evan D.~Skillman\altaffilmark{1},
Julianne J.~Dalcanton\altaffilmark{2},
John M.~Cannon\altaffilmark{3},
Andrew E.~Dolphin\altaffilmark{4},
Jon Holtzman\altaffilmark{5},
Daniel R.~Weisz\altaffilmark{2}, and
Benjamin F.~Williams\altaffilmark{2}
}

\altaffiltext{1}{Minnesota Institute for Astrophysics, School of Physics and
Astronomy, 116 Church Street, S.E., University of Minnesota,
Minneapolis, MN 55455, \ {\it kmcquinn@astro.umn.edu}} 
\altaffiltext{2}{Department of Astronomy, Box 351580, University 
of Washington, Seattle, WA 98195}
\altaffiltext{3}{Department of Physics and Astronomy, 
Macalester College, 1600 Grand Avenue, Saint Paul, MN 55105}
\altaffiltext{4}{Raytheon Company, 1151 E. Hermans Road, Tucson, AZ, 85756}
\altaffiltext{5}{Department of Astronomy, New Mexico State University, Box 30001-Department 4500, 1320 Frenger Street, Las Cruces, NM 88003}

\begin{abstract}
We map the spatial distribution of recent star formation over a few$\times$100 Myr timescales in fifteen starburst dwarf galaxies using the location of young blue helium burning stars identified from optically resolved stellar populations in archival \textit{Hubble Space Telescope} observations. By comparing the star formation histories from both the high surface brightness central regions and the diffuse outer regions, we measure the degree to which the star formation has been centrally concentrated during the galaxies' starbursts, using three different metrics for the spatial concentration. We find that the galaxies span a full range in spatial concentration, from highly centralized to broadly distributed star formation. Since most starbursts have historically been identified by relatively short timescale star formation tracers (e.g., H$\alpha$ emission), there could be a strong bias towards classifying only those galaxies with recent, centralized star formation as starbursts, while missing starbursts that are spatially distributed. 
\end{abstract} 

\keywords{galaxies:\ starburst -- galaxies:\ dwarf -- galaxies:\ evolution -- galaxies:\ individual (ESO~154-023, UGC~4483, UGC~9128, NGC~625, UGC~784, NGC~1569, NGC~2366, NGC~4068, NGC~4163, NGC~4449, NGC~5253, NGC~6456, NGC~6789, IC~4662, DDO~165)}

\section{Introduction}\label{intro}
\subsection{The Starburst Mode of Star Formation in Dwarf Galaxies }
The term starburst refers to an intense, finite period of star-formation (SF) in a galaxy \citep[e.g.,][]{Gallagher1993, Heckman1997, Kennicutt2005}. The increased rate of ongoing SF in a starburst has numerous observational signatures. The youngest, most massive stars produce significant enhanced emission at ultraviolet (UV), optical, and infrared wavelengths, which can be used to identify regions of significant, ongoing SF. However, if these stars are exclusively used in the analysis, their inherently short lifetimes limit identification of extended SF events. Indeed, a starburst can be both spatially and temporally larger than the present-day population of young, high-mass (M $>15$ \msun) stars may suggest. For example, sites of significant, ongoing SF have been identified in numerous galaxies based on their H$\alpha$ emission \citep[e.g.,][]{Lee2009}, but the timescale for H$\alpha$ emission is $\simeq 5-10$ Myr. Thus, the presence of H$\alpha$ can neither confirm nor rule out whether the SF has persisted on longer timescales or has been present in regions other than where the present-day massive stars are located. Moreover, SFRs measured from H$\alpha$ emission are prone both to stochastic effects from the low number of high-mass stars responsible for the emission, and to increased uncertainties due to the emission being a secondary process of the photoionization of gas in low density media (i.e., not from the stars themselves) \citep{Lee2011}. SFRs measured from UV emission are found to be more robust, as the UV flux is emitted directly from the stellar photospheres of more numerous, intermediate mass (2 \msun\ $\ltsimeq$ M $\ltsimeq15$ \msun) B-stars \citep{Lee2011}. UV emission has the added advantage of sampling a longer timescale than H$\alpha$ emission ($\sim100$ Myr versus $\sim5-10$ Myr) and has been shown to be an excellent tracer of SF in star forming galaxies \citep{GildePaz2007}. 

Compared to UV and H$\alpha$ emission, optical imaging of resolved stellar populations can sample higher time resolutions and much longer timescales (depending on the photometric depth achieved by the observations). Using high-fidelity photometry of resolved stellar populations, the longer temporal baselines available at optical wavelengths can be used to reconstruct the history of SF over Gyr-based timescales \citep[for a review, see][and references therein]{Gallart2005} providing a much broader perspective of the SF experienced in a host galaxy. 

Among the many morphological types in which starbursts have been seen, dwarf galaxies offer valuable laboratories in which to study the starburst phenomenon. Dwarf galaxies have the advantage of being both the most numerous morphological galaxy type (i.e., providing a larger possible sample of local starbursts within a fixed volume), and having low metal content (i.e., and thus less internal extinction than higher metallicity systems simplifying the analysis of the star formation histories (SFHs)). The solid-body rotation of dwarf galaxies \citep[e.g.,][]{Skillman1988, Skillman1996} prevents the mixing of the stellar populations on short timescales, which facilitates spatial analysis of the recent SF \citep[e.g.,][]{Gieles2008, Bastian2009, Bastian2011}. Starbursts in dwarf galaxies are known to create outflows of chemically enriched material into the intergalactic medium \citep{Martin2002, Summers2003, Hartwell2004}. This outflow of gas from a starburst has been linked to both potentially playing a role in transforming dwarf irregular galaxies into dwarf spheroidal or dwarf elliptical systems \citep[e.g.,][]{Marlowe1997, Mayer2001, vanZee2001, Pasetto2003} and in shaping the magnetic field of the host galaxy \citep[e.g., NGC~1569;][]{Kepley2010}. Finally, dwarf galaxies have low stellar surface densities, which makes them easily resolved into stars, facilitating analysis of their CMDs.

\subsection{Spatially Centralized vs. Distributed Starbursts}
Historically, starbursts have been assumed to be centralized. This assumption is based on observations showing that recent SF is often found primarily in the center of a starburst galaxy's optical disk. This effect is seen in different morphological types, including starburst spiral galaxies \citep[e.g.,][]{Caldwell1999}, irregular starburst dwarf galaxies, and blue compact dwarf galaxies \citep[BCDs; a sub-class of starburst dwarf galaxies, e.g.,][]{Papaderos1996, vanZee1998, Salzer1999}. Theoretical simulations of starburst galaxies also often assume a central starburst \citep[e.g.,][]{Norman1988, Mihos1994, MacLow1999}. 

However, there is ample evidence that starbursts may not be a solely centralized phenomenon. For example, \citet{Meurer2000} studied the UBI colors of star clusters and diffuse light in the starburst spiral galaxy NGC~3310 and found that 80\% of the flux in the U band was emission from the diffuse SF throughout the optical disk of a galaxy and only 20\% came from the centrally concentrated stellar clusters. A similar study reported both centrally and non-centrally concentrated SF in nearby BCDs based on the radial profiles of $B-V$ colors \citep{Hunter2006}. Distributed starbursts have been reported in both low and high stellar density regions in individual galaxies based on resolved young stars \citep[e.g.,][]{Annibali2003, Cannon2003}. In addition, observations have been better fit by simulations with widespread, shock-induced bursting SF, rather than centrally concentrated, density-dependent SF \citep{Chien2010}. The metrics used to evaluate the concentration of SF in these heterogeneous studies vary, preventing a direct, uniform comparison of results. However, the mixed findings emphasize that one cannot assume that starbursts are solely a centralized phenomenon.

\subsection{Re-Examining the Starburst Paradigm}
This paper is the third in a series studying starbursts in nearby dwarf galaxies. In the first paper, \citet[][hereafter, Paper I]{McQuinn2010a}, we derived the star formation histories (SFHs) of twenty galaxies by matching the optically resolved stellar populations of the galaxies to synthetic color-magnitude diagrams (CMDs). These individual SFHs allow us to define a starburst galaxy as a system with unusually high levels of SF compared to its own historical average \citep{Kennicutt1998a}. This perspective widens the definition to include lower-intensity starbursts that have neither the immediate extreme observational signatures of the youngest most massive stars nor the large super-star clusters / strong emission produced by high absolute SFRs. Using this broader definition, four galaxies in the sample were newly identified as recent starbursts (i.e., Antlia dwarf, ESO~154$-$023, NGC~784, and UGC~9128). Indeed, the optical images of these bursts do not show the larger super star clusters and H$\alpha$ emission typically associated with well-known starburst galaxies such as NGC~5253, NGC~4449, or NGC~1569 (which are also included in our sample). Yet, the SFHs of these newly identified starbursts clearly show elevated recent star formation rates (SFRs) and qualify the systems as starbursts according to the criterion in Paper~I (i.e., recent SFRs~$> 2 \times \langle$SFR$\rangle_{historical}$). Five additional galaxies were identified as fossil starbursts that have present-day SFRs decreasing to historical levels or below. 

In the second paper of the series, \citet[][hereafter, Paper II]{McQuinn2010b}, we focused on the temporal behavior of the starbursts. The starburst durations were measured to last a minimum of 450 Myr in nineteen of the twenty dwarf galaxies, and as much as 1.3 Gyr in two of the systems. SFRs sustained on these timescales cannot be supported by an individual pocket of SF, due both to the limited availability of gas and to the likely disruption of SF by feedback on small spatial scales. Thus, these longer durations suggest that a starburst can be more than individual pockets of flickering SF occurring in clusters and that a significant fraction of a galaxy may participate in a burst. Yet, it is unclear how much of the galaxy experiences a burst and whether the SF occurs in all regions simultaneously or propagates across a galaxy. 

In this third paper of the series, we study the spatial distribution of SF in fifteen of the original twenty starburst dwarf galaxies with deep, wide area \textit{Hubble Space Telescope} (HST) coverage. The goal of the paper is to test the assumption that starbursts are generally a centralized phenomenon. To accomplish this goal, we trace the location of SF over the last few 100 Myr in each galaxy using positions of young, intermediate mass (2 \msun\ $\ltsimeq$ M $\ltsimeq 15$ \msun) blue core helium burning (BHeB) stars from archival \textit{HST} observations. In addition, we reconstruct the temporally resolved SFHs for regions of varying stellar density, thus tracking the SFR across different regions of each galaxy. 

The paper is organized as follows$\colon$ \S \ref{obs} summarizes the observations and data processing. \S\ref{spatial_analysis} presents the metrics used to determine whether the starbursts are centralized or distributed and discusses the results of these metrics. \S\ref{conclusions} summarizes the conclusions.

\section{The Galaxy Sample, Stellar Populations, and Regional SFHs}\label{obs}

\subsection{The Galaxy Sample and CMDs}\label{sample}
The sample of twenty starburst dwarf galaxies studied in Papers~I and II was selected from observations in the $HST$ public archive. The observations were chosen based on three properties that ensure robust reconstruction of the recent SFHs. First, we selected galaxies that lie close enough (i.e., D $\ltsimeq 5$ Mpc) to resolve their stellar populations with $HST's$ imaging instruments $-$ the Advanced Camera for Surveys \citep[ACS;][]{Ford1998} or the Wide Field Planetary Camera 2 \citep[WFPC2;][]{Holtzman1995}. Second, we required the archival observations to include both V and I band images of a galaxy. Third, we selected systems where the I band observations' photometric depth reached a minimum of $\sim2$ mag below the tip of the red giant branch $-$ a requirement for accurately constraining the recent SFH of the galaxy \citep{Dolphin2002, Dohm-Palmer2002b}. One galaxy (SBS~1415$+$437) did not meet this final requirement but was included in the original studies for comparison purposes. It is not analyzed in this paper. While our archival-defined sample spans a wide parameter space for starburst dwarf galaxies including many of both the stronger, well-known starburst dwarf galaxies in the nearby universe (e.g., NGC~1569, NGC~4449, NGC~5253) and the smaller, less well-known starbursts (e.g., NGC~4068, UGC~9128), it is not a volume limited sample. 

From this group of nineteen galaxies, we applied two additional cuts to ensure global spatial analysis was possible. First, we required that the $HST$ observations covered at least 50\% of the deprojected area out to the B-band 25 mag arcsec$^{-2}$ isophotal diameter (D$_{25}$), as defined by \citet{Karachentsev2004}. This criterion eliminated three galaxies with limited areal $HST$ coverage (NGC 4214, NGC 6822, Holmberg~II). Second, we required that the number of core helium burning stars allowed for a meaningful spatial analysis, which required that the amplitude of the burst must be greater than $10^{-3}$ \msun~yr$^{-1}$. This criterion eliminated one galaxy (Antlia) from the sample. The final sample consists of the fifteen dwarf galaxies listed in Table~\ref{tab:properties}. Also included in Table~\ref{tab:properties} is the percentage of the deprojected D$_{25}$ area covered by $HST$ and the equivalent area in kpc$^2$. The deprojected area was calculated by correcting the angular area of each galaxy for the optical inclination angle (\textit{i}), Galactic extinction in the B-band, and the distance to the system. The specific $HST$ instruments, observation times, and $HST$ program IDs can be found in Paper~I (Table~1), and \citet{Weisz2008}.

\begin{figure}
\begin{center}
\includegraphics[width=\columnwidth, clip=true]{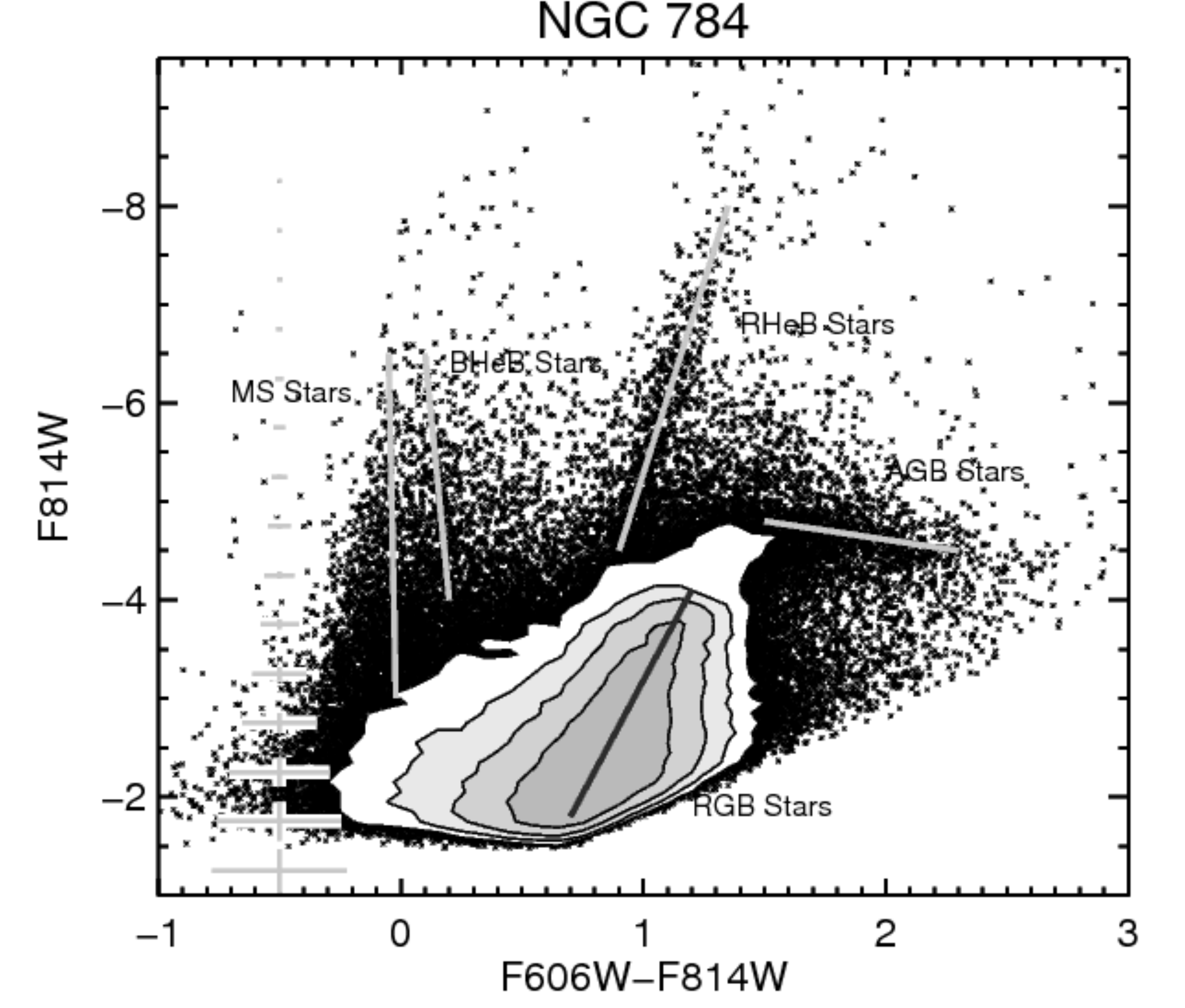}
\caption{The CMD of NGC~784 showing the typical photometric quality of the data used in the analysis. Photometric uncertainties are shown in the left hand side of the CMDs. Major stellar evolutionary stages are marked on the CMD. Contours are used to show the photometric structure of the stellar populations in the red giant branch and lower main sequence.}
\label{fig:example_cmd}
\end{center}
\end{figure}

All photometry was measured by DOLPHOT or HSTphot \citep{Dolphin2000} from the standard HST pipeline processed and cleaned images. The data sets were processed uniformly, ensuring reliable comparison between galaxies. Artificial star tests were performed to measure completeness using the same photometric software packages listed above. The details of the photometric processing for fourteen of these galaxies are presented in Paper I and for one galaxy (DDO~165) in \citet{Weisz2008}. CMDs for all galaxies can be found in Figure~2, Paper~I. Here in Figure~\ref{fig:example_cmd}, we present an example CMD for NGC~784, whose data reach our minimum required photometric depth. Representative uncertainties are shown per magnitude by crosses in the left hand side of the CMD. 

Stellar evolutionary features that are readily apparent in the example CMD of Figure~\ref{fig:example_cmd} include the main sequence (MS), blue and red core helium burning (BHeB and RHeB respectively) sequences, the red giant branch (RGB), and the asymptotic giant branch (AGB). The upper MS, the BHeB, and the RHeB stars are populated by stars $\gtsimeq$ 2 \msun\ and thus are unambiguous indicators of recent star formation \citep[e.g.,][and references therein]{Gallart2005, McQuinn2011}. Stars with a mass of 2 \msun\ or greater will evolve off the MS and begin burning helium in their cores. These RHeB stars form a sequence slightly hotter than the RGB stars. At brighter magnitudes, the RHeB stars are well-isolated from other populations, but at fainter magnitudes they begin to become indistinguishable from the RGB stars. The RHeB stars remain at cooler effective temperatures until their opacity changes due to internal convective mixing. These stars then become hotter and rapidly traverse the ``blue loop'' in the CMD becoming BHeB stars, located just redward of the upper MS stars in CMD space \citep[e.g.,][]{Stothers1991}. When the core helium is exhausted, the stars will evolve further into either AGB stars, moving redward in the CMD to cooler temperatures, or, erupt as supernovae. 

\subsection{HeB Stars Trace the Locations of Recent SF}\label{heb_tracers}

As described in \S\ref{sample} and seen in Figure~\ref{fig:example_cmd}, the BHeB and RHeB stars occupy a unique space in the CMD. Along the RHeB and BHeB sequences, there is a nearly one-to-one correspondence between the luminosity of a HeB star and the mass of the star. The masses and stellar ages of each point on the sequences can be estimated using stellar evolution isochrones (see \S\ref{dating_heb}). Thus, HeB stars in a given luminosity range can be robustly age-dated using stellar evolution isochrones \citep[e.g.,][]{Dohm-Palmer1997}, and the spatial distribution of HeB stars of different ages can be used to trace the sites of recent SF. Note that the luminosity of a HeB star does undergo small changes as the star evolves. This minor change in luminosity is smaller than our photometric uncertainties and we do not correct for it. Here we present our methodology for isolating HeB stars, dating them using stellar evolution libraries, and thus tracing the sites of recent SF. We use one galaxy (NGC~784) as an example. In the Appendix, we present the results of this analysis for the entire galaxy sample.

\subsubsection{Isolating the HeB Stars in an Optical CMD}\label{isolate_heb}

At low metallicities, both blue and red HeB stars are relatively isolated in an optical CMD as seen in Figure~\ref{fig:example_cmd}. The bright BHeB stars form a sequence that is approximately parallel to and slightly redward of the upper MS; the exact shape of the BHeB sequence depends on the host galaxy's metallicity, and to a lesser extent, the SFH. At fainter magnitudes, the sequence of BHeB stars trends redder and eventually blends into the red clump. The typical intrinsic V$-$I color separation between the MS and BHeB stars is $0.1-0.3$ mag. This distinction can become blurred in galaxies with higher metallicities, internal reddening, and photometric crowding. However, the starburst galaxies in our sample have low metallicity and low internal reddening. They also have less stellar photometric crowding than larger spiral galaxies on average, although dwarf starburst galaxies do have regions of higher stellar density. Overall, photometric uncertainties and crowding are relatively small and do not significantly blur the MS and BHeB populations, as can be seen in the representative CMD in Figure~\ref{fig:example_cmd}. The separation between the populations is more apparent at higher luminosities, and becomes less distinct at lower luminosities depending upon the photometric uncertainties. There are three exceptions in our sample (NGC~1569, NGC~4449 and NGC~5253), where the MS and BHeB populations begin to blend due to photometric crowding and differential extinction. Nonetheless, we include these three galaxies in the analysis noting higher uncertainties in the spatial analysis of BHeB stellar population. Note also the RHeB stars form a sequence blue-ward of the RGB stars; these stars blend with the RGB stars at luminosities below the tip of the red giant branch (TRGB) making them difficult to uniquely separate in a CMD for ages $\gtsimeq100$ Myr for most galaxies \citep{McQuinn2011}.

The technique of separating the MS, BHeB, and RHeB stars is described fully in \citet{Dohm-Palmer2002a} and \citet{McQuinn2011}. While the MS and BHeB sequences are readily identifiable in the CMDs, separating these sequences by eye can be ambiguous, particularly at lower luminosities (see Figure~\ref{fig:example_cmd} as an example). To robustly separate the MS and BHeB stars, we created histograms of the V$-$I colors of all stars in the MS, BHeB and RHeB regions of the CMD, in intervals of 0.5 to 1.0 magnitudes in M$_V$, depending on the photometric uncertainties and on the number of HeB stars present. Binned in this way, the concentration of stars forming the MS, BHeB, and RHeB sequences form three distinct peaks in the stellar color distribution. These three histogram peaks mark the center position in the CMD of each stellar sequence. We conservatively selected the stars with colors of $\pm0.1$ mag of the peaks for both the blue and red HeB populations in each galaxy. 

\begin{figure}
\begin{center}
\includegraphics[width=\columnwidth, clip=true]{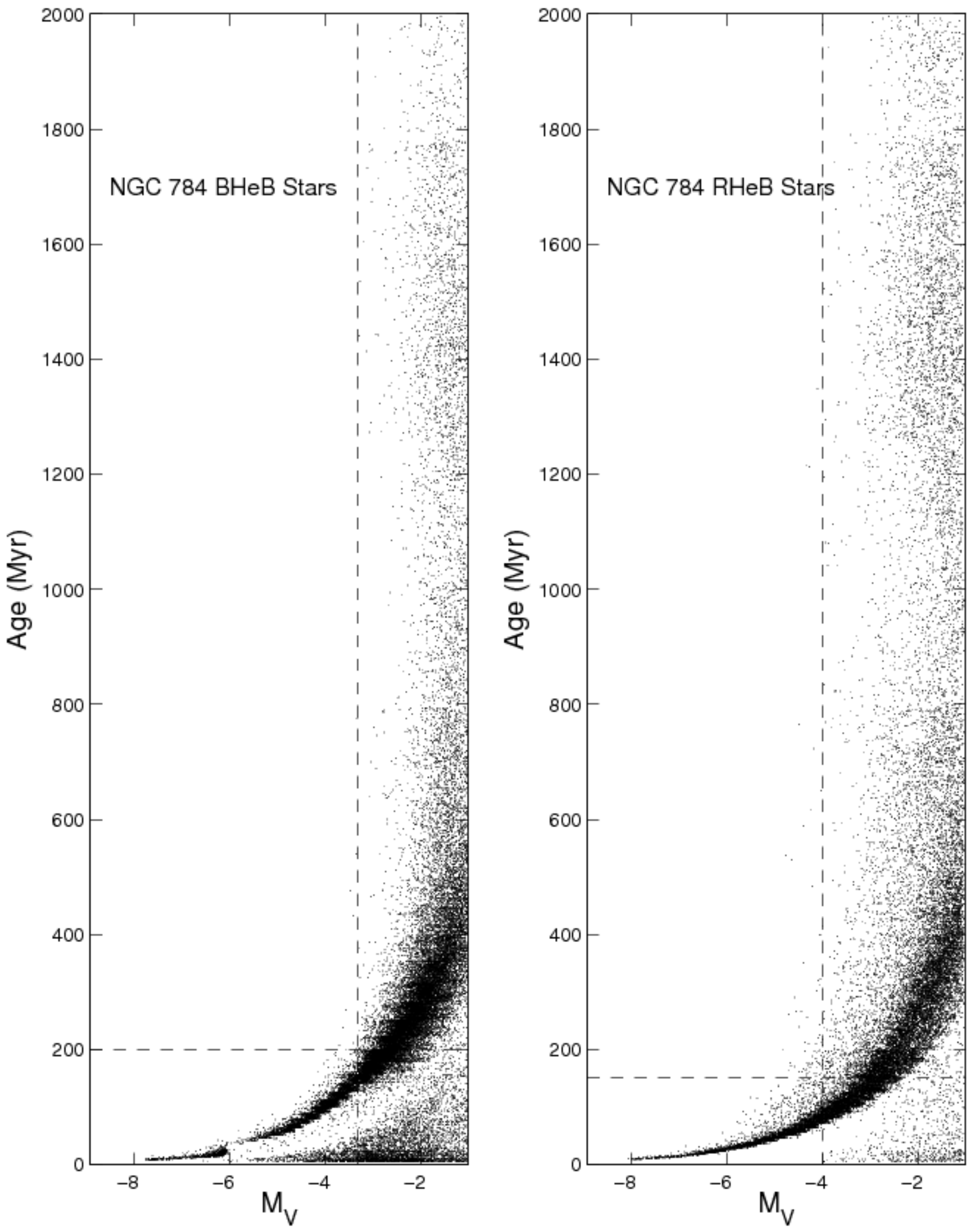}
\caption{The age distribution of stars in the blue (left panel) and red (right panel) HeB stellar regions as a function of magnitude for a simulated CMD matching the completeness of the data for NGC~784. There is a one-to-one correspondence between age and luminosity for the HeB stars until the HeB stars begin to mix with red clump or RGB stars at fainter magnitudes. This luminosity is a lower limit for unambiguously selecting HeB stars from other stellar populations in a CMD. For this galaxy, the limit corresponds to an age of $\sim200$ Myr for BHeB stars and $\sim150$ Myr for the RHeB stars. The limit is unique for each system and dependent upon photometric uncertainties, crowding, metallicity, and the SFH of the host galaxy.}
\label{fig:heb_age}
\end{center}
\end{figure}

At fainter magnitudes, the BHeB and RHeB stars begin to merge with the red clump and RGB stars respectively, creating a lower luminosity limit for selecting HeB stars from a CMD. This limit is specific to each data set, because it depends on the metallicity, photometric uncertainties, and SFH of each galaxy. The individual lower limits were determined by creating synthetic CMDs from the SFHs (Paper~I) and by mimicking photometric completeness and observational uncertainties using artificial star tests performed on the original data sets. In the left panel of Figure~\ref{fig:heb_age}, we plot age versus luminosity for the stars in the BHeB region of a synthetic CMD, based on the best fit SFH of NGC~784. The one-to-one age-luminosity relation of the BHeB stars is apparent in Figure~\ref{fig:heb_age}, until an age of $\sim200$ Myr, where there are stars of multiple ages present for a given luminosity. It is at this magnitude where the BHeB sequence for the simulated CMD of NGC~784 becomes mixed with older red clump stars. BHeB stars cannot be unambiguously selected at magnitudes fainter than this limit. Likewise, as seen in the right panel of Figure~\ref{fig:heb_age}, the RHeB region becomes contaminated with older RGB stars at low luminosities, corresponding to an age limit of $\sim150$ Myr. In general, we find that the BHeB stars can be traced further back in time than the RHeB stars, making them a longer range chronometer of recent SF.

\subsubsection{Dating the HeB Stars using Theoretical Isochrones}\label{dating_heb}

Once the blue and red HeB stars were identified in the CMD, we used stellar evolution isochrones to age date the stars. Stellar evolution isochrones track the paths of stars of different mass in a CMD.  We use the theoretical isochrones of \citet{Girardi2000} for stars with masses $\leq 7$ \msun, with improved treatment of the asymptotic giant branch (AGB) stars added by \citet{Marigo2008}. At higher masses (M $>$7 \msun), we use the isochrones of \citet{Bertelli1994}. This set of isochrones, collectively known as the Padova stellar evolution isochrones, is used to age-date the HeB in the observed CMDs. Isochrones were matched in metallicity and adjusted for both the internal and foreground extinction for each galaxy. The individual metallicities were taken from \HII\ region abundance measurements in the literature whenever possible. In five cases, the metallicity was estimated from the luminosity-metallicity relation for dwarf galaxies \citep[i.e,][]{Zaritsky1994,Tremonti2004,Lee2006}. The $A_V$ extinction measurements were estimated from the SFHs (Paper~I) and transformed to $A_I$ extinction using the York Extinction Solver tool \citep{McCall2004}, assuming the \citet{Cardelli1989} reddening law with R$=3.1$. These values are in good agreement with the values from the dust maps of \citet{Schlegel1998}, but also incorporate estimates of internal extinction. The adopted metallicity and extinction values can be found in \citet[][Table~1]{McQuinn2011}. We opted not to use the isochrones incorporating circumstellar dust models \citep{Bressan1998, Groenewegen2006}, as they primarily affect the AGB stage of evolution and do not affect the location of HeB stars in the CMDs.
 
\begin{figure}
\begin{center}
\includegraphics[width=\columnwidth, clip=true]{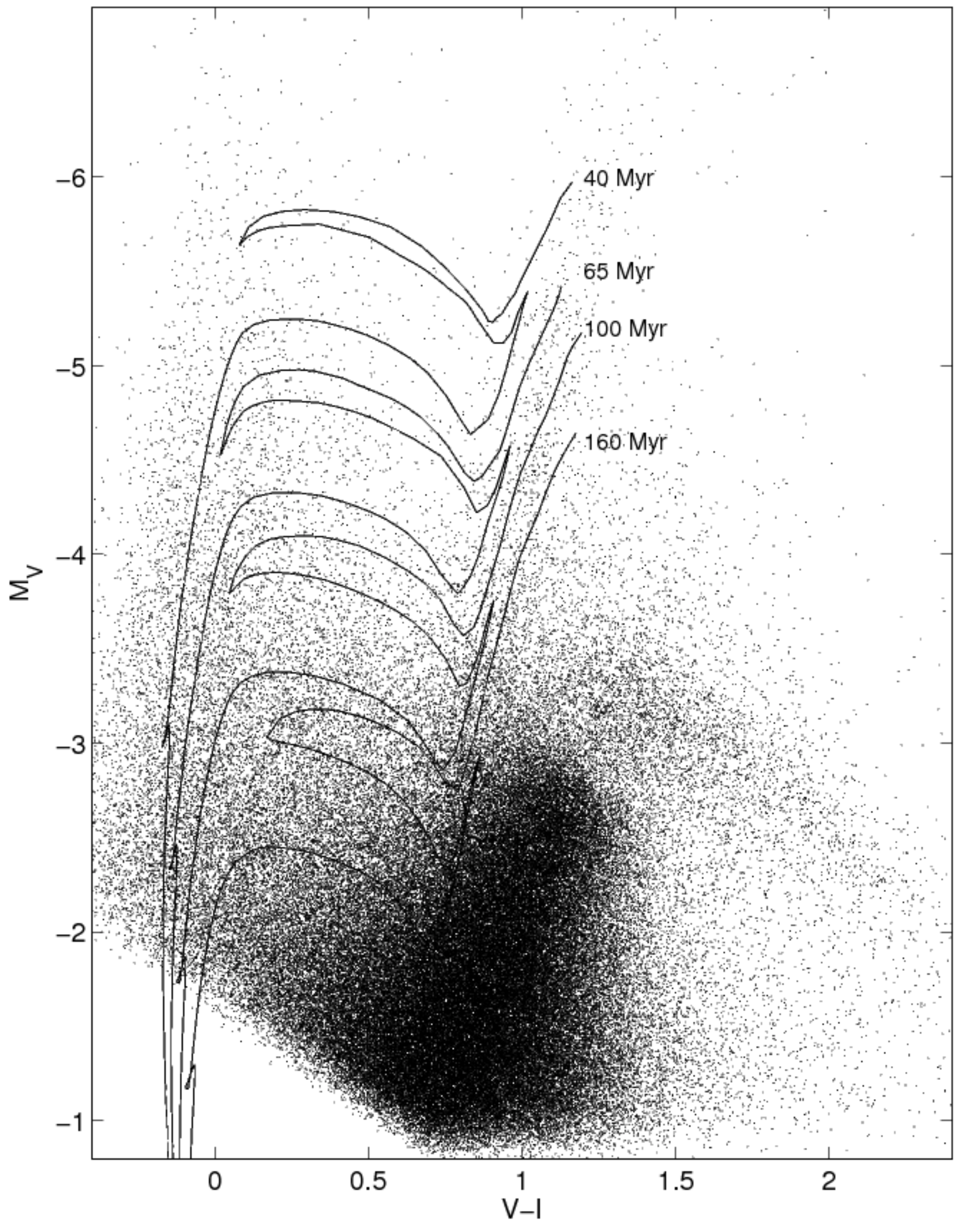}
\caption{The CMD for NGC~784 overlaid with stellar evolution isochrones from 40 Myr to 160 Myr in increments of $log~\delta$$t \equiv 0.2$. The blue and red HeB stages of evolution, easily identified in the isochrones, are in general agreement with the corresponding sequences of HeB stars in the observed CMD.}
\label{fig:cmd_isochrones}
\end{center}
\end{figure}

In Figure~\ref{fig:cmd_isochrones}, we present a representative comparison between the the stellar populations for NGC~784 and the stellar evolution isochrones up to 160 Myr. The blue and red HeB stages of evolution are easily identified in the isochrones, as stars in these stages occupy extreme blue and red turning points in the isochrones. The isochrones are in general agreement with the MS locus and the blue and red extremes of the blue loop sequence.

\begin{figure}
\begin{center}
\includegraphics[width=\columnwidth, clip=true]{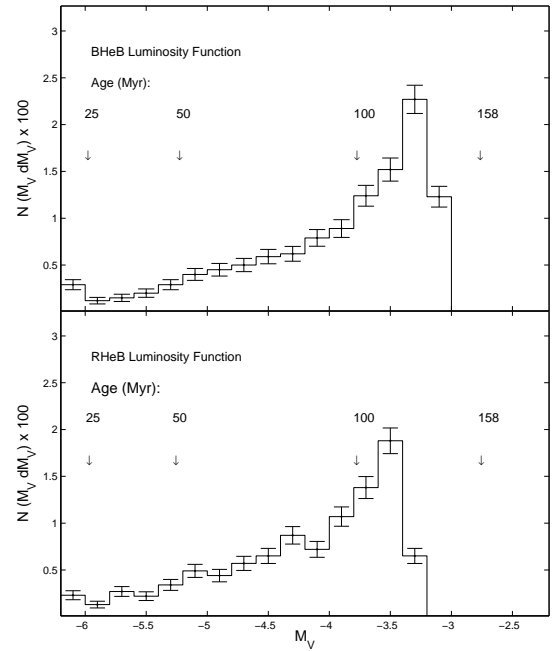}
\vspace{-0.5in}
\caption{The luminosity functions of the BHeB (top panel) and RHeB (bottom panel) stars for NGC~784. Listed above the histogram functions are the ages of the stars derived from Padova stellar evolutionary models. The luminosity functions are plotted only to the fainter magnitudes below which the HeB stars begin to merge with other stellar populations (i.e., M$_V \sim -3.5$ for the BHeB stars and M$_V \sim-3$ for the RHeB stars). The BHeB and RHeB luminosity functions are similar in profile, as expected. Their spatial distributions, shown in Figure~\ref{fig:example_image_heb}, are also very similar. Thus, we focus on only the BHeB stars for the remainder of the analysis. }
\label{fig:lum_fcn}
\end{center}
\end{figure}

In Figure~\ref{fig:lum_fcn}, we plot the luminosity function of the BHeB and RHeB stars selected from the CMD for NGC~784, along with the ages derived from the stellar evolutionary isochrones. As expected, there are smaller numbers of short-lived high mass stars. Uncertainties in the luminosity function include both Poisson errors and incompleteness measurements. The luminosity functions extend to just above the lower luminosity limit where the HeB sequences begin to merge with other stellar populations in the CMD. The luminosity function of the RHeB stars is similar to that of the BHeB stars providing a consistency check on our technique.

\subsubsection{Mapping the Distribution of HeB Stars at Different Ages}\label{map_ages}
We tag each bin in the luminosity function with an equivalent age, determined from the stellar isochrones. Representative ages are shown as arrows in Figure~\ref{fig:lum_fcn}. The locations of the HeB stars in each age bin can then be mapped directly onto the optical image of the galaxy. We chose to map the HeB stars at four different luminosity (or age) bins. The bin selection was different for each system and was based on the structure of each observed CMD, the number of HeB stars present, and the luminosity at which the HeB stars merged into other stellar populations (see \S\ref{isolate_heb} and \S\ref{dating_heb}). 

\begin{figure}
\begin{center}
\includegraphics[width=\columnwidth, clip=true]{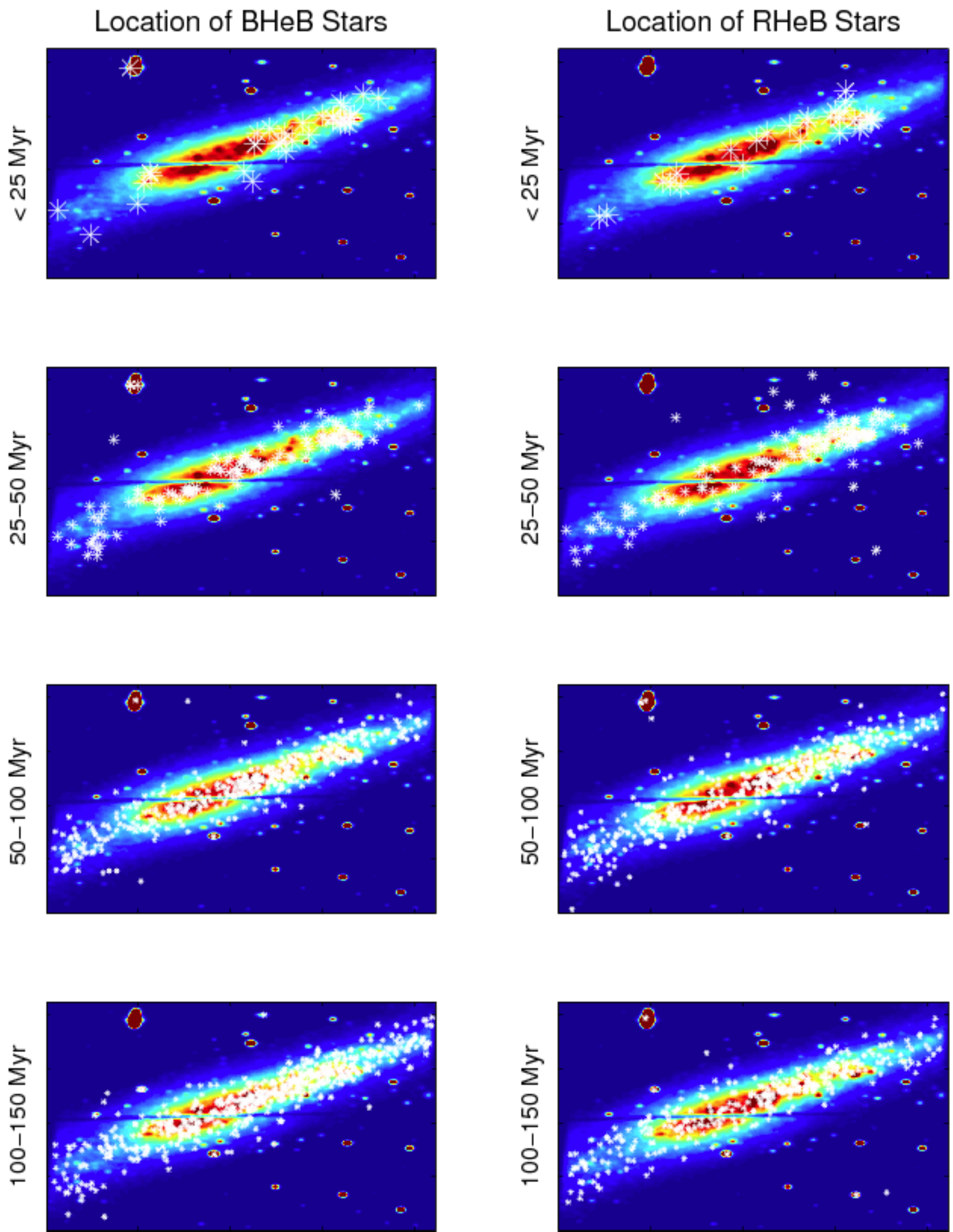}
\caption{The location of the BHeB stars (left panels) and RHeB stars (right panels) of different ages in NGC~784 overlaid on a smoothed V band HST image. The areas of the asterisks are scaled by the average SFR in that age bin to account for the IMF and stellar lifetimes so that the total area in the asterisks is proportional to the SFR in that age bin. Thus, while the number of points in a given age bin is determined by the number of stars, the total area of the points represents the SFR after normalization by the IMF. This allows the simultaneous presentation of the location of the SF, the total SFR, and the number of stars from which the SFR is computed. Similarly to the luminosity functions in Figure~\ref{fig:lum_fcn}, the distribution and number of RHeB stars is in good agreement with the BHeB stars. We present only the distribution of the BHeB stellar population for the remainder of the paper as they can be uniquely separated in a CMD farther back in time.}
\label{fig:example_image_heb}
\end{center}
\end{figure}

In Figure~\ref{fig:example_image_heb}, we plot the positions of the BHeB and RHeB stars on the optical image of NGC~784 for each of four age bins. The background image in each panel is a smoothed V band image made by applying a 25 pixel (i.e., 1.25\arcsec) radius Gaussian kernel to the original HST observation. Included in Figures~\ref{fig:lum_fcn} and \ref{fig:example_image_heb} are the results for both the blue and red HeB stars in NGC~784. Both the luminosity functions and spatial patterns in the two stellar populations are similar, as expected. Thus, we focus on the BHeB population for the rest of the paper as these stars are both more numerous than the RHeB stars \citep[e.g.,][]{Dohm-Palmer2002a, McQuinn2011}, and can be traced to fainter magnitudes in the observed CMDs (i.e., older ages). 

For each age bin, the HeB stars are plotted as white asterisks in Figure~\ref{fig:example_image_heb}. The areas of the asterisks are scaled by the average SFR in that age bin to account for the IMF and stellar lifetimes so that the total area in the asterisks are proportional to the SFR in that age bin. Thus, while the number of points in a given age bin is determined by the number of stars, the total area of the points represents the SFR after normalization by the IMF. This allows the simultaneous presentation of the location of the SF, the total SFR, and the number of stars from which the SFR is computed. As there will naturally be fewer high mass, younger stars, scaling the asterisks in this way enables a more direct comparison of the amount of SF occurring in each age bin. For example, in Figure~\ref{fig:example_image_heb}, the total area in the asterisks in the most recent $0-25$ Myr compared to those $25-50$ Myr ago indicates that the SFR was higher more recently.

\begin{figure}
\begin{center}
\includegraphics[width=\columnwidth, clip=true]{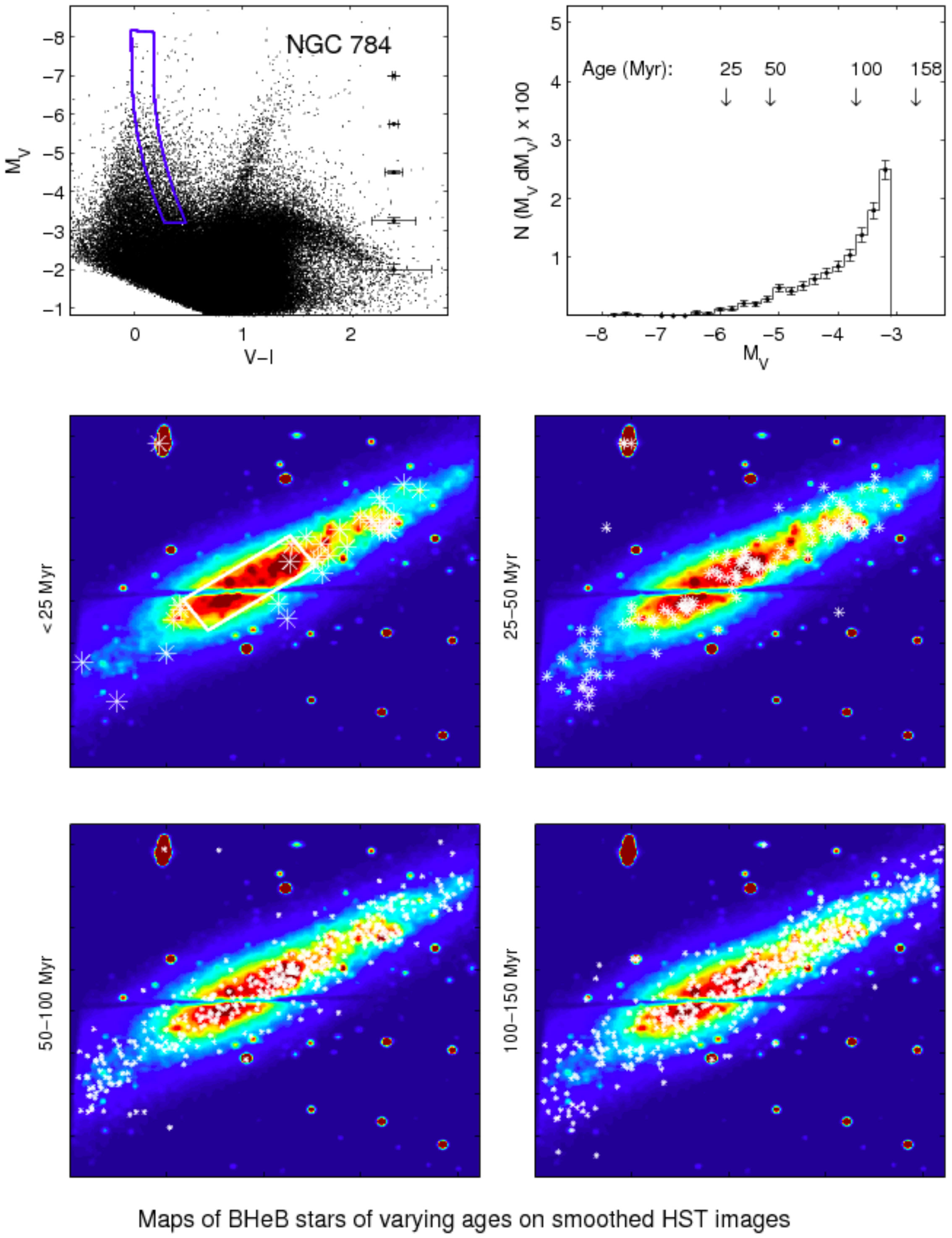}
\caption{Top left panel: CMD of NGC~784 with the region containing BHeB stars used in the analysis marked in blue. Top right panel: The luminosity function of the BHeB stars identified in the CMD. Middle and lower panels: The location of the BHeB stars of four different age ranges overlaid on smoothed V band HST images of the galaxy. The area of the points is scaled by the SFR enabling a direct comparison between age bins. The white box drawn on the first image encompasses the area used in reconstructing the SFH of the central region. The region outside of this box comprises the outer region used in our analysis. Note that the HST images are not shown in WCS orientations. Similar figures for the entire sample can be found in the Appendix.}
\label{fig:example_6panel}
\end{center}
\end{figure}

Combining the above analysis, in Figure~\ref{fig:example_6panel}, we plot the CMD, BHeB luminosity function, and maps of the spatial distributions of the BHeB stars in four age bins for NGC~784. Similar figures mapping the sites of recent SF in each galaxy of the sample can be found in the Appendix. Two of the galaxies, NGC~6789 and NGC~4163, host fossil bursts than ended 80 Myr and 100 Myr ago respectively. Similarly to the other galaxies in our study, we plot the spatial distributions of the BHeB stars in these galaxies in four time bins, but define the first time bin in the maps to be the post-burst time period. We limit our subsequent spatial analysis in these two galaxies to the SF occurring during the active starburst (i.e., prior to this post-burst time period). 

\subsubsection{Inherent Age Limit and Diffusion Timescale of Stars}
Using BHeB stars to trace the sites of recent SF is based on the assumption that the stars have not migrated significantly from their birth sites. Dwarf galaxies are known to be solid-body rotators lacking sheer and differential rotation \citep[e.g.,][]{Skillman1988, Skillman1996}. Their dynamics implies that there should be little radial mixing of stars over timescales considered here. The limiting timescale for our technique is therefore the diffusion timescales for dissolving SF structures. 

Stellar populations within galaxies have been shown to be born with a high degree of substructure made up of a continuous distribution of clusters, groups, and associations \citep[][and references therein]{Bastian2010}. For a sample of dwarf galaxies, \citet{Bastian2011} found these structures persist on the order of hundreds of Myr. This timescale was measured down to the limiting photometric depth of the data and is thus a lower limit. These findings are similar to previously reported lower limit lifetimes of $\sim80$ Myr for stellar structures in the SMC \citep{Gieles2008} and $\sim175$ Myr for the LMC \citep{Bastian2009}. While the stellar structures could not be probed on longer timescales given their data, it is assumed that the stellar structures will continue to dissolve, including some radial migration of the stellar populations. Although our data are frequently sensitive to HeB stars as old as 400 Myr, we limit the spatial analysis to HeB stars younger than 250 Myr, the approximate lower limit found by \citet{Bastian2011} for stellar structure to be maintained in dwarf galaxies. We list the adopted limiting timescale for each galaxy in Table~\ref{tab:properties}. 

\section{Evaluating the Spatial Concentration of the Starbursts}\label{spatial_analysis}
As described in \S\ref{intro}, while starbursts have often been assumed to be centrally concentrated, there is also observational evidence that starbursts can be distributed in nature. The current data set of temporally and spatially resolved stellar populations allow us to measure the spatial extent of the SF. We focus on quantifying this spatial distribution using two timescales. The first is over the last $100-250$ Myr using BHeB stars, and the second is over the duration of each starburst event. In the following subsections, we discuss three techniques employed to evaluate the concentration of bursting SF over these two timescales. 

\subsection{Three Metrics for Measuring the Concentration of SF \label{metrics}}
We employ three techniques to measure the spatial concentration of the SF. The first technique compares the spatial distribution of the young stellar populations as traced by the BHeB stars (ages $\ltsimeq100-250$ Myr) with the underlying older stellar populations as traced by the RGB stars (ages $\gtsimeq1$ Gyr). This metric evaluates the concentration of SF over one timescale corresponding to the oldest BHeB stars (see Table~\ref{tab:properties}, Column 6). The second technique compares the SFHs reconstructed from the central and outer regions. The third technique estimates the fraction of stellar mass created in the central region. Because both the second and the third metrics use the temporally resolved SFHs, they can be evaluated over different timescales. We chose two timescales. The first matches that defined by the BHeB stars used in the first metric. The second is equal to the duration of the burst events (see Table~\ref{tab:properties}, Column 5). 

\subsubsection{First Metric: BHeB Star Concentration}
Our first technique uses the BHeB stars selected in \S\ref{isolate_heb}, shown in Figure~\ref{fig:example_6panel} for NGC~784 and in the Appendix for the entire sample. We compare the distributions of BHeB stars to those of the RGB stars. The RGB stars were selected from the well-populated and well-defined RGB based on magnitude and color cuts. For example, in Figure~\ref{fig:example_6panel}, we selected RGB stars within a region defined by $-3.2<$M$_V<-1$ and $0.8<$V-I$<1.3$. The actual region used is curved and follows the distribution of RGB stars in the CMD. We fit the distributions of the BHeB and RGB stellar number densities with 2-D elliptical Gaussian functions. While the stellar distributions are not perfectly described by a Gaussian profile, a Gaussian function provides a good first order approximation of the distribution. Uncertainties are estimated to be of order 10\% in fitting the full-width at half-maximum (FWHM) of the stellar distributions based on the errors on the fit parameters. One system, NGC~6789, was dropped from this particular analysis due to difficulties fitting low numbers of BHeB stars. The degree of central concentration of the young stars can be measured using the ratio of the FWHM of the two stellar populations' distributions. 

\begin{figure}
\begin{center}
\includegraphics[width=\columnwidth, clip=true]{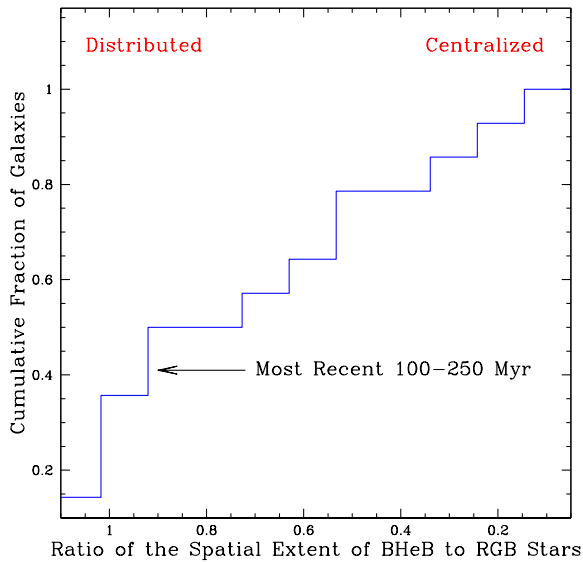}
\vspace{-1in}
\caption{Cumulative distribution of the ratio of the spatial extent of the young BHeB stars ($\ltsimeq100-250$ Myr) versus older RGB stars ($\gtsimeq1$ Gyr). The distribution shows a continuous spectrum of the spatial extent of recent SF in the sample. In some of the galaxies, the FWHM of the BHeB stars spatial distribution is confined to within 25\% of the FWHM of the RGB distribution, indicating highly centralized SF. In other systems, the FWHM of the BHeB stars extends to over 80\% of the RGB distribution, indicating distributed SF. There are also a number of galaxies where the BHeB stars extend to roughly half the extent of the RGB stars, emphasizing that the spatial distribution lies along a continuum.}
\label{fig:bheb_rgb_histo}
\end{center}
\end{figure}

In Figure~\ref{fig:bheb_rgb_histo}, we plot the cumulative distribution of this ratio for the sample. There is a continuous spectrum of the spatial extent of recent SF in these systems. There is no dependence on the depth of BHeB photometry in the distribution of the ratio measurements. In some galaxies, the BHeB stars are confined to within 25\% of the RGB spatial distribution, indicating centralized SF. In other galaxies, the BHeB stars extend to more than 80\% of the RGB spatial distribution, indicating distributed SF. However, there are a number of galaxies where the BHeB stars extend over roughly half the RGB spatial distribution, emphasizing the point that the spatial distribution of SF lies along a continuum. The SF in these galaxies are neither clearly centralized nor clearly distributed, but fall in an intermediate category. We will discuss these results further when comparing the results of the three metrics.

\subsubsection{Second Metric: b Parameter ratios}
Our second technique for measuring the concentration of SF analyzes the SFHs of the central and outer regions of each galaxy. We define the ``central'' regions to be within where the V band surface brightness drops to $10-25$\% of the peak flux in the unsmoothed $HST$ images. In one case, NGC~2366, a large star forming complex lies outside the central region making the flux cut-off between central and outer fall at the upper end of this range. In Figure~\ref{fig:example_6panel}, the central region for our example galaxy NGC~784 is outlined on the smoothed images of the galaxies; the outer regions encompass the remainder of the field of view. We used rectangular regions to define the central regions. We found that our analysis is robust against small changes in the delineation between ``central'' and ``outer'' regions and the shape of the regions, because of the large number of stars. The central and outer regions for the entire galaxy sample are shown in the Appendix.

We reconstructed separate SFHs for the central and outer regions, using the same technique and methodology that we applied to the whole field of view in Paper~I. Briefly, we derived the star formation rate ($SFR(t,Z)$) applying a CMD fitting technique \citep{Dolphin2002} with stellar evolutionary models from \citet{Marigo2007} to the photometry and artificial star lists in the central and outer regions. When necessary, additional artificial stars were generated to adequately sample the completeness in small areas. We fit for distance, extinction, and stellar age, and constrain the metallicity to increase with time (except in NGC~2366 where the photometry reached a full magnitude below the red clump). The fit distances were in good agreement with TRGB measured distances from the literature. We refer the reader to Paper I for a full discussion of the SFHs, the time binning used, and uncertainties. 

In Figure~\ref{fig:regional_sfhs}, we present the SFHs of the last 1.5 Gyr for both the central and outer regions in the galaxy sample. The CMD fitting technique used to reconstruct these SFHs uses information from all the stellar populations in the CMDs, not just the HeB stars discussed in the preceding analysis. Thus, the SFHs are not limited by the same constraints in selecting blue or red HeB stars and can extend over a much longer timeline. A horizontal bar shows the burst durations for the each galaxy as measured in Paper~I. 

\begin{figure}
\begin{center}
\includegraphics[width=\columnwidth, clip=true]{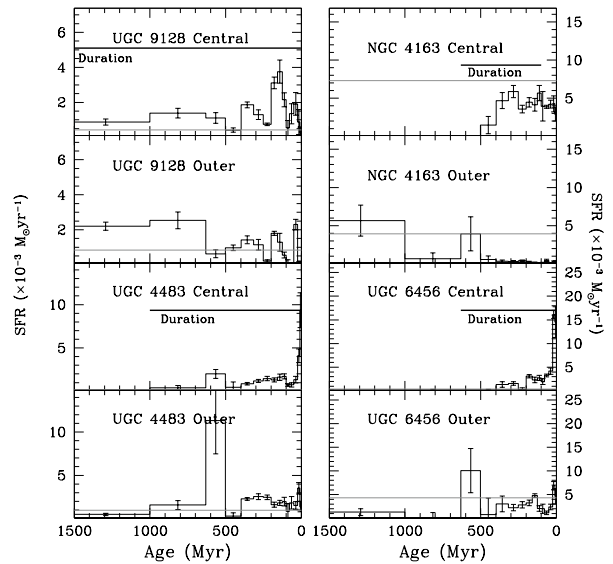}
\vspace{-1in}
\caption{The SFHs of regions of the central and outer regions of the galaxies over the past 1500 Myr. The central regions are defined for each galaxy in Figure~\ref{fig:galaxies} in the Appendix. The duration of the starbursts from (Paper~II) are drawn as a bar at the top of each plot. The lifetime average SFR (i.e., birthrate parameter$=1$) derived from each region is drawn as a solid gray line.}
\label{fig:regional_sfhs}
\end{center}
\end{figure}

A useful metric to quantify starbursts is the birthrate parameter, defined as the ratio of the current SFR to the past average \citep[b~$\equiv$~SFR~/~$\langle$SFR$\rangle$,][]{Scalo1986,Kennicutt1998a}. Starbursts have been found to have birthrate parameters greater than 2; for comparison, current birthrate parameters for normal star-forming galaxies are typically of order 0.5 \citep{Kennicutt2005}. In Figure~\ref{fig:regional_sfhs}, the solid horizontal line designates a value b$ = 1$ for the individual region.

\begin{figure}
\begin{center}
\figurenum{\ref{fig:regional_sfhs}}
\includegraphics[width=\columnwidth, clip=true]{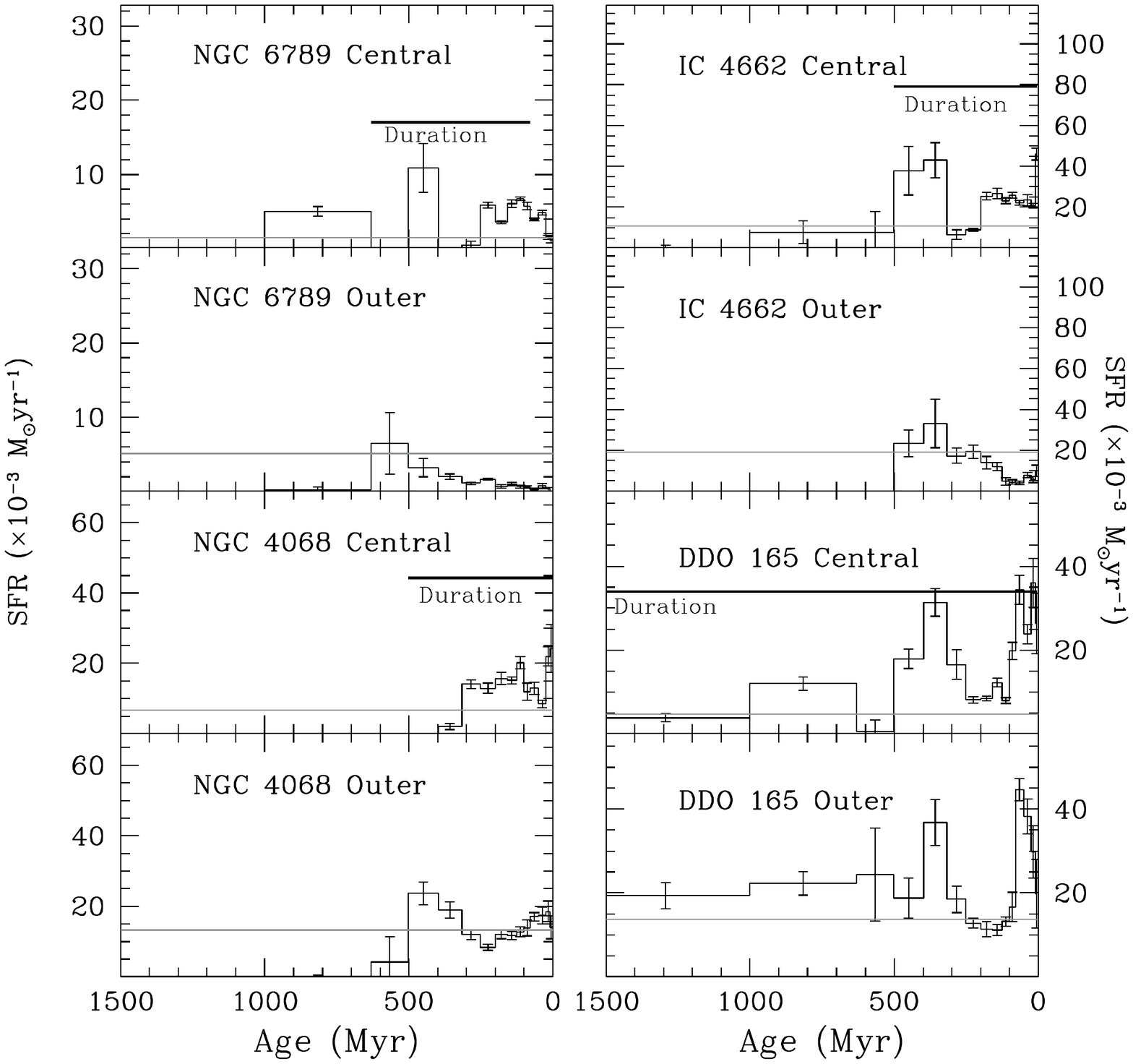}
\vspace{-1in}
\caption{\textit{Cont.}}
\end{center}
\end{figure}

\begin{figure}
\begin{center}
\figurenum{\ref{fig:regional_sfhs}}
\includegraphics[width=\columnwidth, clip=true]{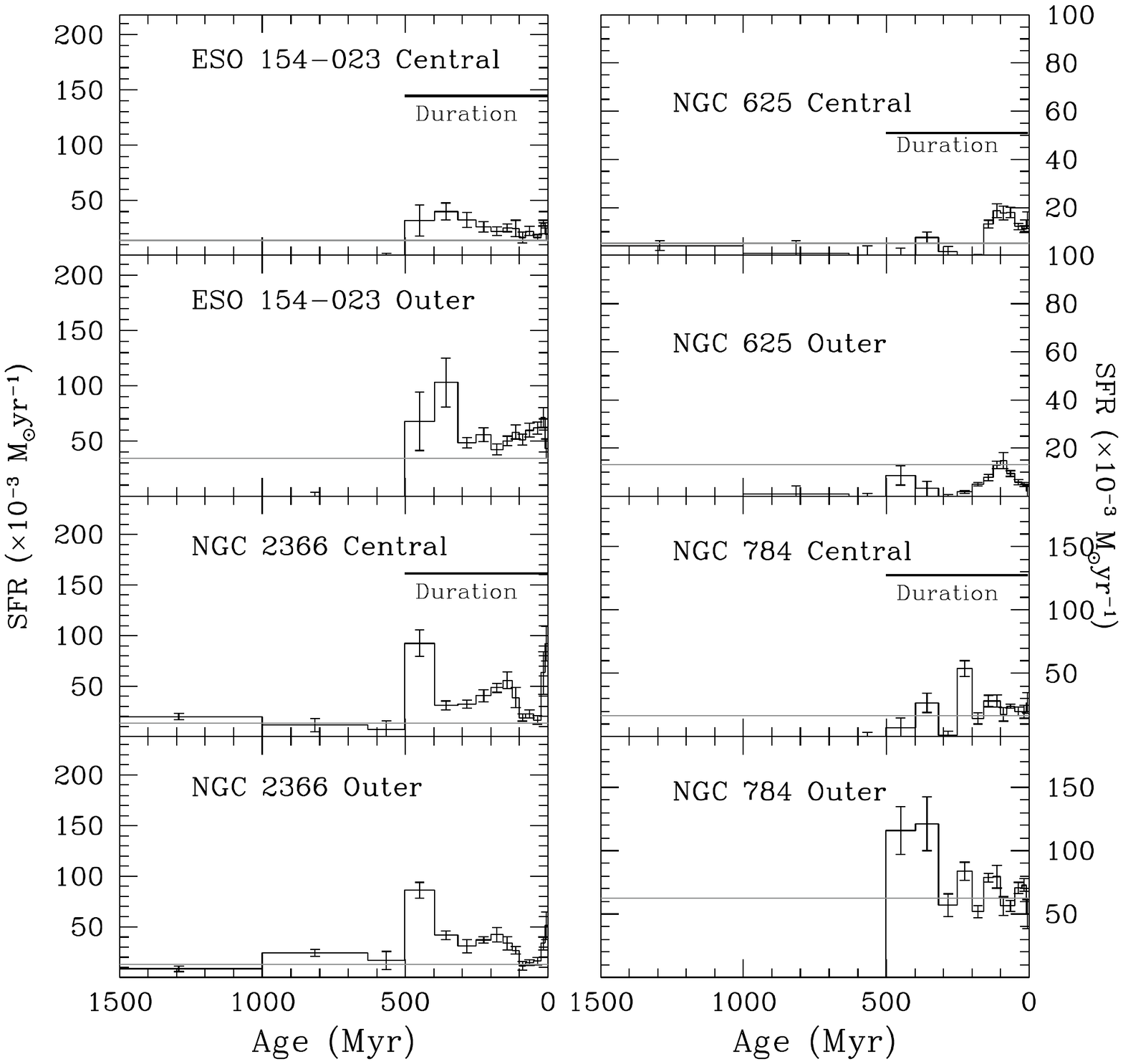}
\vspace{-1in}
\caption{\textit{Cont.}}
\end{center}
\end{figure}

\begin{figure}
\begin{center}
\figurenum{\ref{fig:regional_sfhs}}
\includegraphics[width=\columnwidth, clip=true]{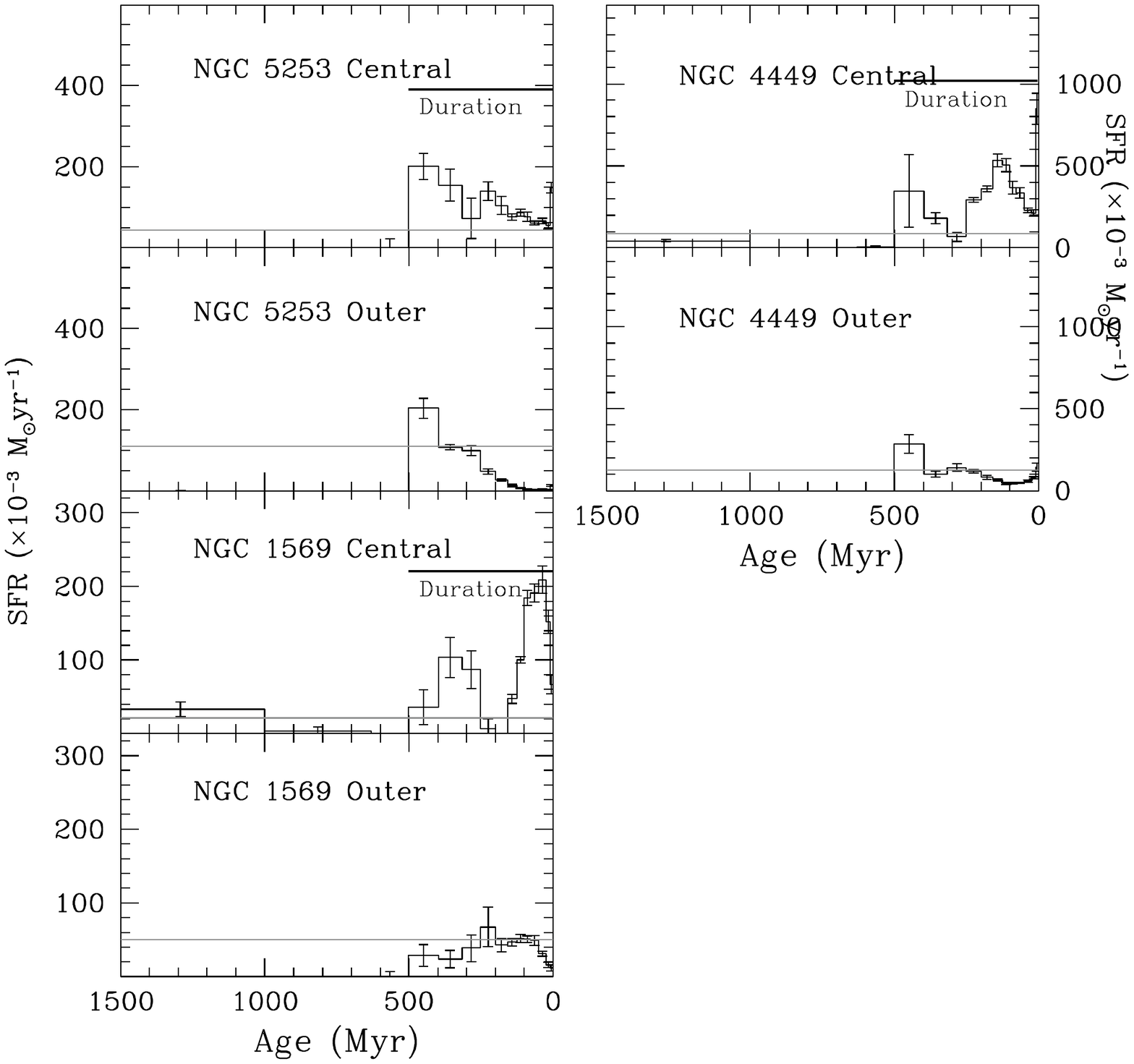}
\vspace{-1in}
\caption{\textit{Cont.}}
\end{center}
\end{figure}

\begin{figure}
\begin{center}
\includegraphics[width=\columnwidth, clip=true]{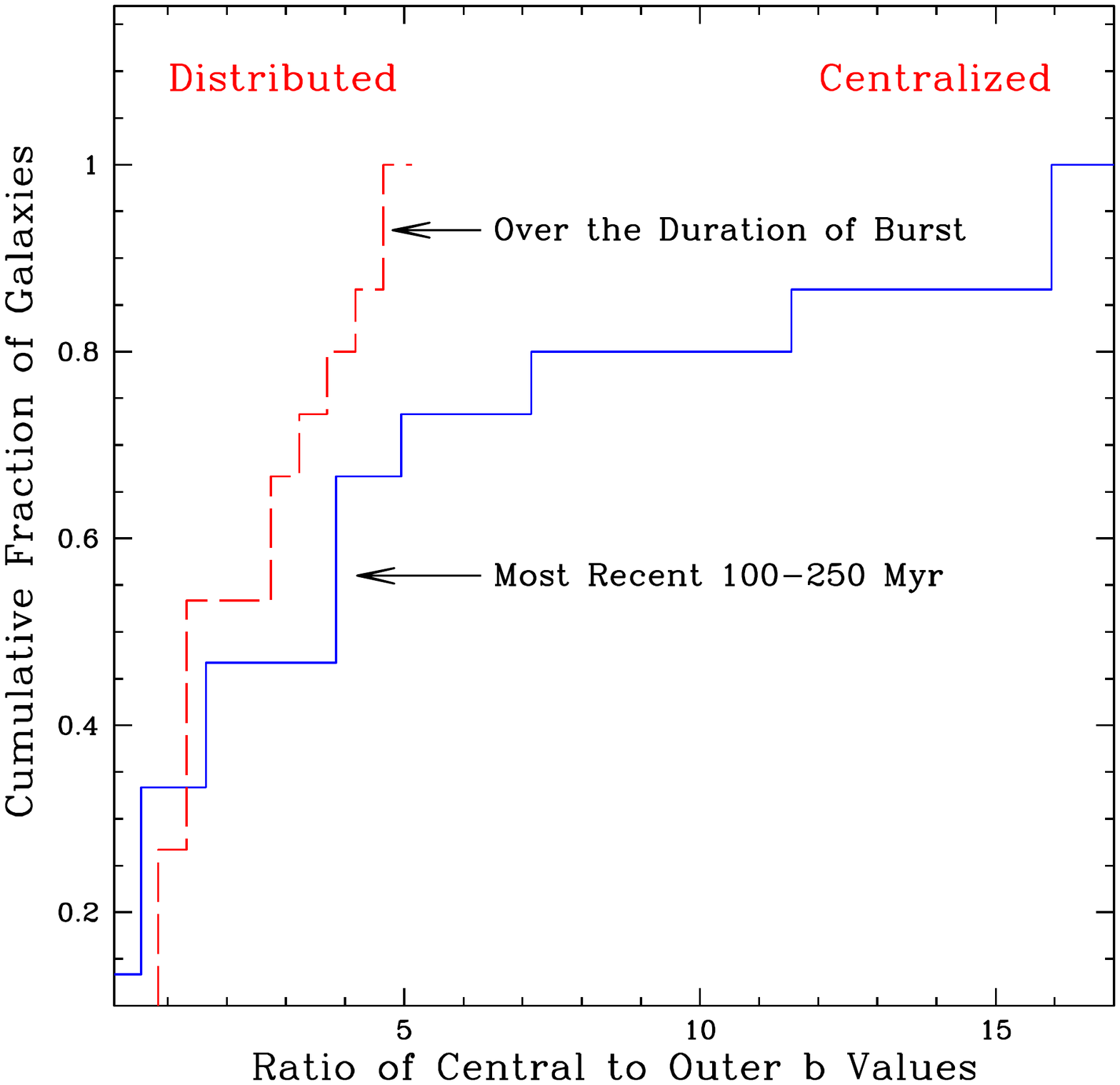}
\vspace{-1in}
\caption{Cumulative distribution of the ratios of birthrate parameters in the central to outer regions evaluated over two timescales. The ratio at more recent times (solid blue line) shows a larger range in the distribution. While a number of galaxies have high ratio values indicating highly centralized SF, nearly half have birthrate parameters in the outer regions comparable to within a factor of two of the values in the central regions, indicating greater spatial distribution in the SF. Not all starbursts are highly centralized. Over the longer timescale of the burst durations (dashed red line), the ratios are smaller overall. This smaller range is partly due to the differences in the time binning used in the most recent $100-250$ Myr and that used over the longer duration timescales. However, despite this bias, there is still a significant range in the spatial distribution of SF in these galaxies when measured over the duration of the starbursts.}
\label{fig:bvalues}
\end{center}
\end{figure}

We are interested in assessing whether the SF in the central regions has been more enhanced than the SF in the outer regions. Thus, we compare the ratio of the birthrate parameter in the central and outer regions for the sample. In Figure~\ref{fig:bvalues}, we plot the cumulative distribution of this ratio, for the two timescales we selected. In the most recent $100-250$ Myr (solid blue line), some of the galaxies have ratios greater than 10, indicating the central regions showed more enhanced SF than the outer regions (i.e., highly centralized SF). However, nearly half of the galaxies have birthrate parameters in their outer regions within a factor of two of the central regions. This indicates that both the central and the outer regions experienced enhanced SF activity. This reinforces our result from the first metric that shows a continuous spectrum of the spatial extent of recent SF. On the longer timescale of the burst durations (dashed red line), the distribution in the ratios is much smaller. This tighter range in the ratio of the birthrate parameter is partly due to the differences in the time binning used in the most recent $100-250$ Myr and that used over the longer duration timescales, as averaging the birthrate parameter over longer timescales eliminates peak values. However, despite this bias, there is still a significant range in the spatial distribution of SF in these galaxies when measured over the duration of the starbursts. 

\subsubsection{Third Metric: Fraction of Stellar Mass Formed}
Our third technique estimates the fraction of the stellar mass formed in the central region. The stellar mass was estimated by integrating the SFRs in the central and outer regions over the same two timescales used in comparing the central and outer SFHs. Note that this third metric is not strictly independent of the second metric as they both use the SFHs to measure the spatial concentration of SF. However, the ratio of the birthrate parameter in the second metric compares how intense the SF is in the inner and outer regions relative to their historical average, whereas the third metric integrates the SFH only over a specific time period. In other words, these two metrics both use the SFHs but have different normalizations and thus provide different measurements of the spatial concentration of SF.

\begin{figure}
\begin{center}
\includegraphics[width=\columnwidth, clip=true]{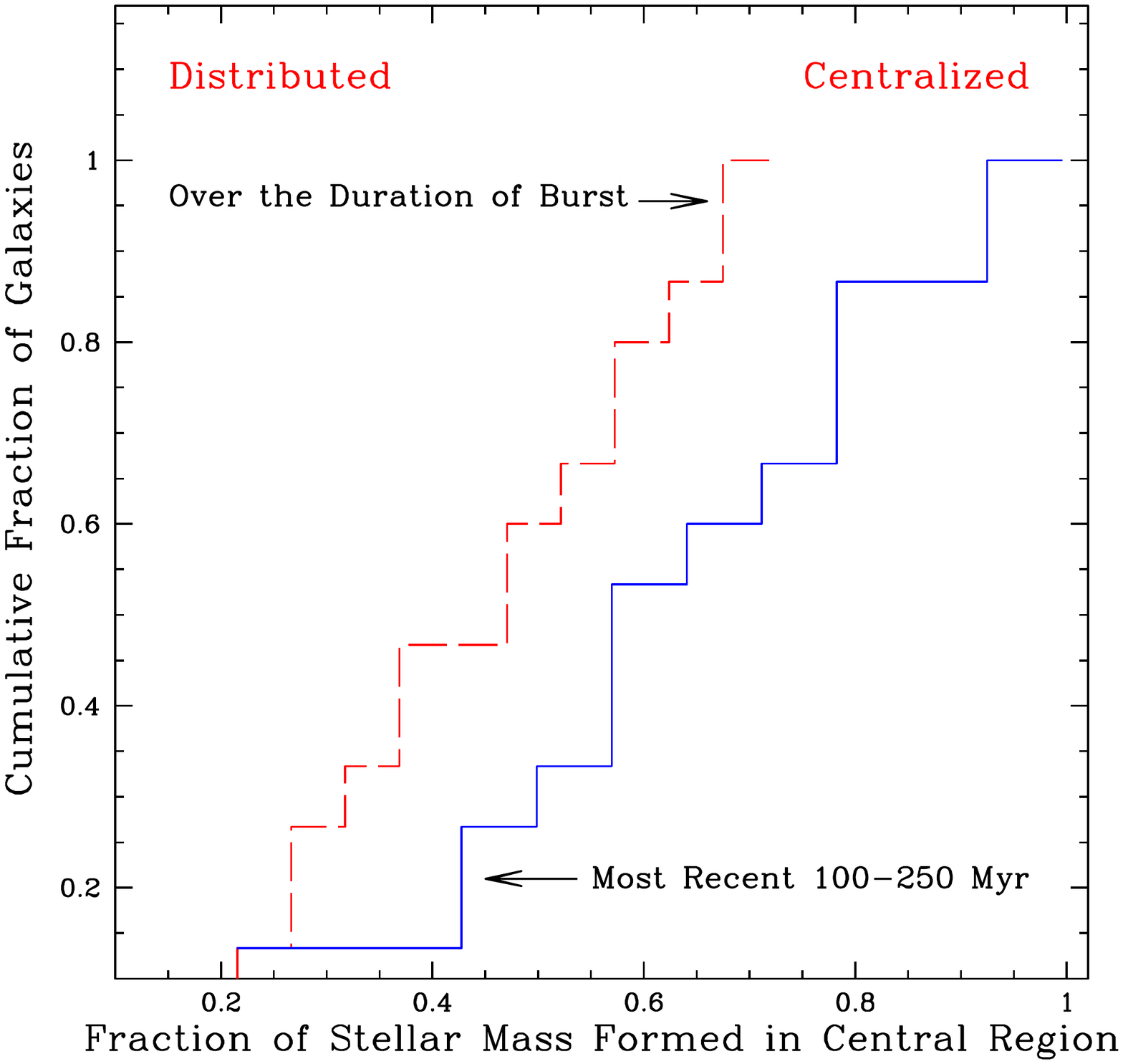}
\vspace{-1in}
\caption{Cumulative distribution of the fraction of stellar mass formed in the central regions of the starbursts over the most recent $100-250$ Myr (solid blue line) and over the lifetime of the burst (dashed red line). Similar to the other two metrics, a continuous distribution is seen in the figure, particularly at more recent times. In some galaxies, over 80\% of the stellar mass is formed in the central regions, indicating centralized SF. In two systems, only 25\% of the stellar mass is formed in the central regions, indicating distributed SF. Over the duration of the bursts, there are more galaxies with a greater percentage of their stellar mass formed in the outer regions.}
\label{fig:stellar_mass_histo}
\end{center}
\end{figure}

In Figure~\ref{fig:stellar_mass_histo}, we present the cumulative distribution of the fraction of stellar mass formed in the central regions. Similarly to the other two metrics, in the most recent $100-250$ Myr (solid blue line) the fraction of stellar mass formed in the central regions shows a continuum of values. In some galaxies, more than 80\% of the stellar mass was formed in the central regions, indicating highly centralized SF. In other systems, less than 25\% of the stellar mass was formed in the central regions, indicating distributed SF. However, there are also a number of galaxies where roughly half of the stellar mass was formed in the central regions, emphasizing the point that the spatial distribution of SF lies along a continuum. When measured over the durations of the bursts (dashed red line), the distribution of the ratio of central to outer stellar mass in Figure~\ref{fig:stellar_mass_histo} is narrower. No galaxy is forming more than 80\% of their mass in the central region, indicating that the SF is less centralized over this longer timescale. 

\subsection{Are the Bursts Centralized or Distributed: A Comparison of Three Metrics}
\begin{figure}
\begin{center}
\includegraphics[width=\columnwidth, clip=true]{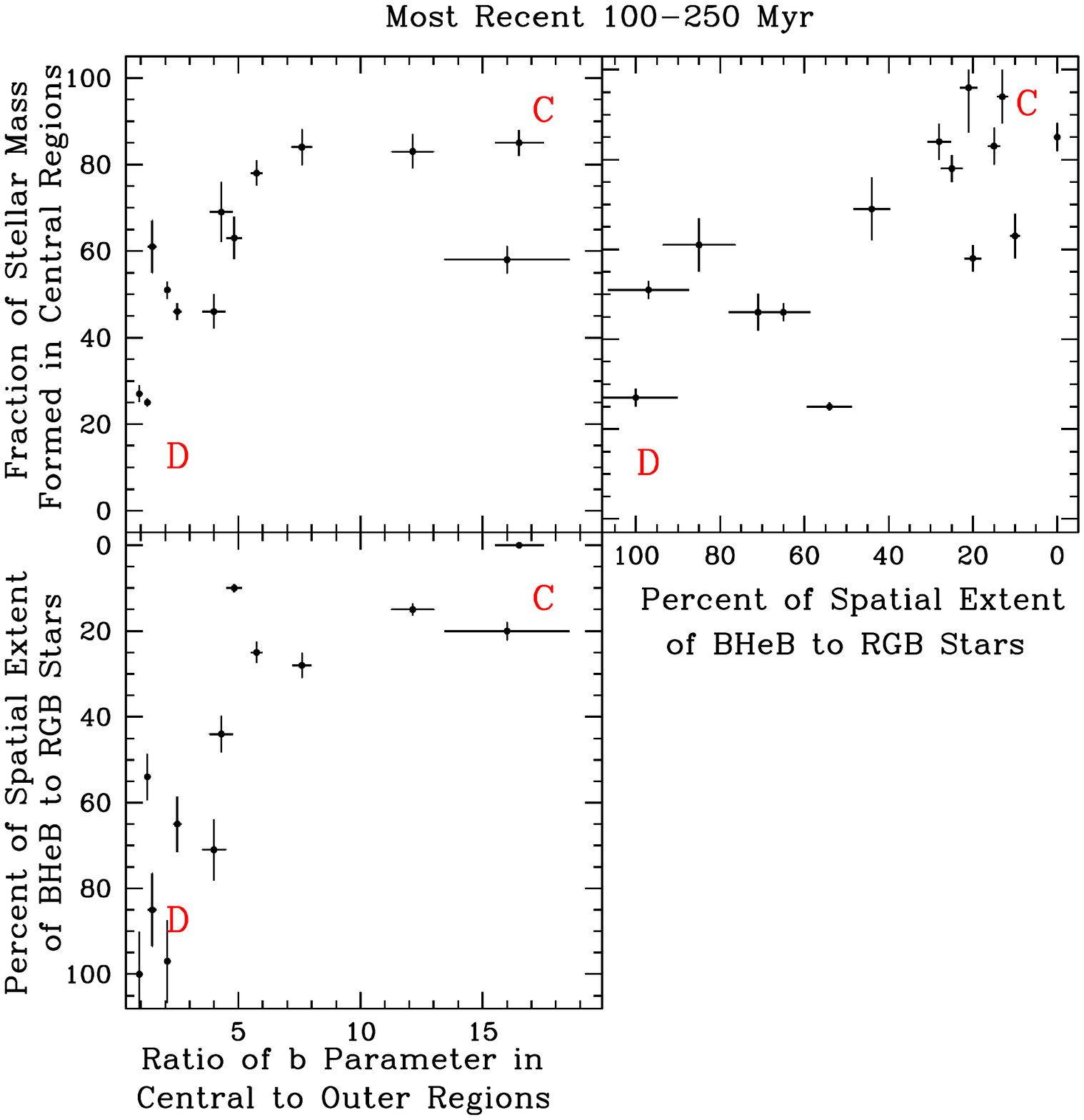}
\vspace{-1in}
\caption{Comparison of three metrics evaluating the spatial distribution of SF over the most recent $\sim100-250$ Myr (see Table~\ref{tab:properties} for time bins on the individual galaxies). The red ``C'' in each panel marks the centrally concentrated end of the spectrum, while the red ``D'' marks the spatially distributed end. The three metrics are correlated and show a continuous spectrum in the spatial extent of SF in these starburst galaxies.}
\label{fig:metrics_recent}
\end{center}
\end{figure}

In Figure~\ref{fig:metrics_recent}, we compare the three metrics for evaluating the concentration of SF over the most recent $100-250$ Myr.  We mark the highly centralized SF end of the continuous spectrum with a red ``C'', and the distributed SF end with a red ``D''. The results for individual galaxies in each metric are listed in Table~\ref{tab:spatial_extent}. Two galaxies (NGC~4163 and NGC~5253) had no detectable SF occurring in the outer regions over this timescale and are not plotted. Based on the SF occurring in the central regions in these two galaxies, the ratio of the BHeB to RGB stars spatial extent were 21\% and 14\%, indicating highly centralized SF.

As seen in Figure~\ref{fig:metrics_recent}, the three metrics are correlated. The systems with much smaller distributions of the younger BHeB stars relative to the older RGB stars show lower birthrate parameters in the outer regions of the galaxies relative to the central regions. These systems also have a larger fraction of stellar mass being formed in the central, high surface brightness regions. Likewise, the galaxies where the spatial distributions of younger BHeB stars are roughly equivalent to the spatial extent of the older RGB stars also show birthrate parameters in the outer regions that are roughly comparable to the central region values. Additionally, smaller fractions of the stellar mass are formed in the central regions in these galaxies. 

As seen in Figure~\ref{fig:metrics_recent}, there is a continuum in the spatial extent of the starbursts. Some galaxies, such as NGC~4163, NGC~5253, and NGC~6789, do show highly centralized SF in the most recent $100-250$ Myr. In contrast, other galaxies, such as NGC~4068, ESO~154$-$023, and NGC~784, show indications of distributed SF. These findings are confirmed in the maps of the BHeB stellar distributions in the Appendix which show that sites of recent SF are highly centralized in some galaxies but can also extend over much of the optical disk of a galaxy in others. The majority of the systems lie somewhere between these two extremes. While delineating between centralized and distributed SF would be based on arbitrary thresholds, it is clear that not all starbursts are highly centralized in the most recent $100-250$ Myr.

\begin{figure}
\begin{center}
\includegraphics[width=\columnwidth, clip=true]{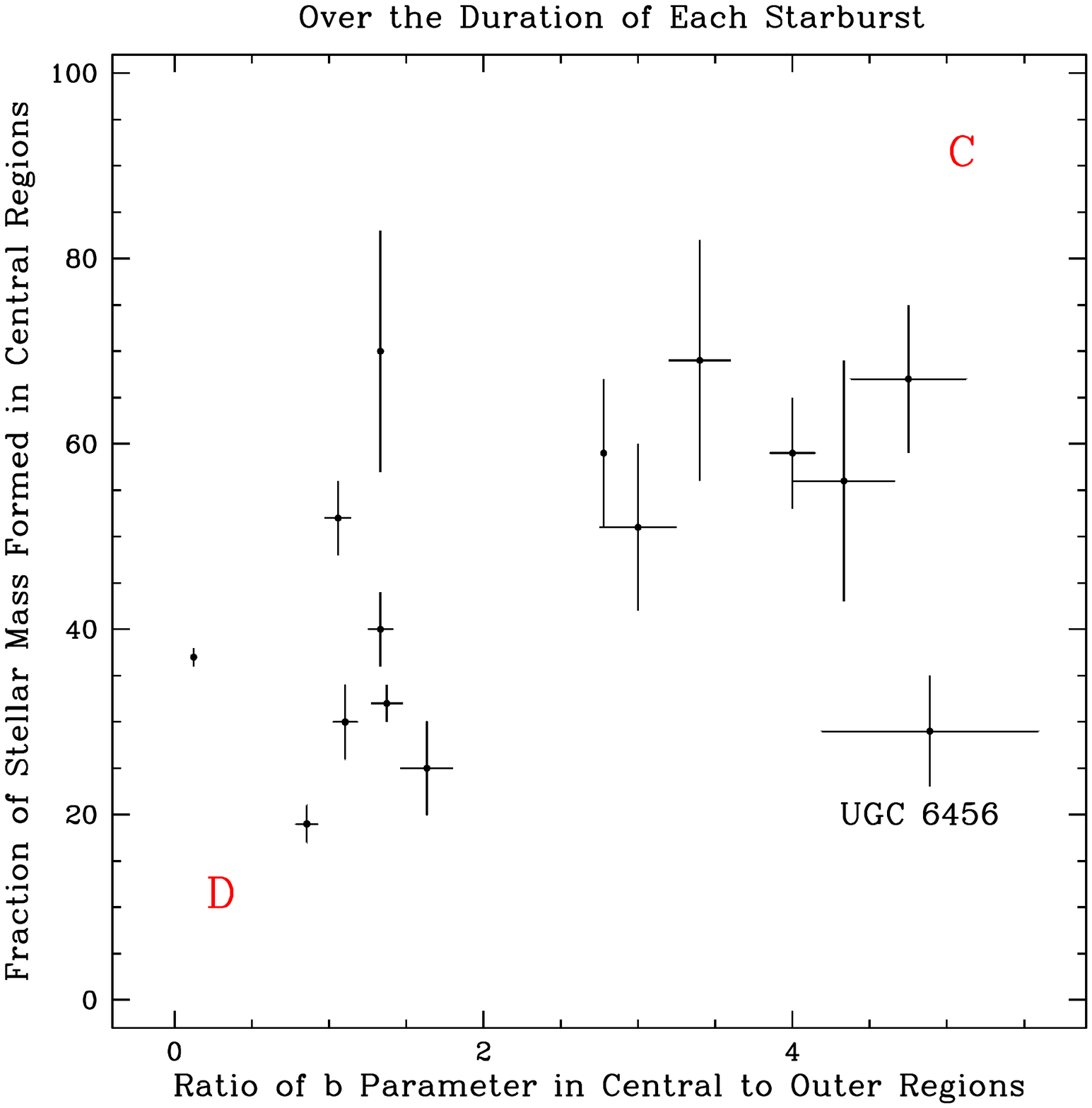}
\vspace{-1in}
\caption{Comparison of two metrics evaluating the spatial distribution of SF over the duration of each starburst. Similarly to Figure~\ref{fig:metrics_recent}, the red ``C'' in each panel marks the centrally concentrated end of the spectrum, while the red ``D'' marks the spatially distributed end. Note the x-axis has a shorter range than the companion plot in Figure~\ref{fig:metrics_recent}. There are number of galaxies that lie in the lower left quadrant indicating that SF was more distributed over these longer timescales. Comparing these results with Figure~\ref{fig:metrics_recent}, we find a range in the spatial extent of the starbursts over both timescales studied.}
\label{fig:metrics_lifetime}
\end{center}
\end{figure}

In Figure~\ref{fig:metrics_lifetime}, we present a comparison of two concentration metrics calculated over the duration of the burst events, which is longer than the $100-250$ Myr timescale in all of the galaxies. Note the plotted range for the ratio in birthrate parameters is smaller in Figure~\ref{fig:metrics_lifetime} than in the companion plot in Figure~\ref{fig:metrics_recent}. There is a general correlation between the two metrics. When the birthrate parameter in the outer regions is within a factor of two of the central regions, the fraction of stellar mass formed in the central regions is also smaller. One galaxy, UGC~6456, is an outlier from the trend.

Comparing Figures~\ref{fig:metrics_recent} and \ref{fig:metrics_lifetime}, we find a range in the spatial extent of the starbursts over both timescales studied. This distribution in the spatial extent of the starbursts did not correlate with absolute magnitude, starburst duration, or average SFRs. Note that our sample was selected based the $HST$ observations available in the public archive, as discussed in \S\ref{sample}, and it is therefore not a volume-limited sample. Thus, we are unable to measure the distribution function of the spatial extent of SF in starburst dwarf galaxies as a class of objects. However, our result does have implications for understanding the trigger mechanisms of the starbursts. Whatever physical mechanisms initiate these starbursts must account for this range in spatial properties.

The galaxies that show the highest degree of centralized SF (e.g., NGC~1569, NGC~4449, NGC~5253) are also the most well-known, and have been identified as starbursts according to numerous and diverse observational criteria, including hosting a population of Wolf-Rayet (W-R) stars \citep{Schaerer1999} or super star clusters \citep{Larsen2008}, having threshold infrared emission line ratios \citep{Thornley2000}, or showing significant H$\alpha$ emission \citep{Lee2009}, and high SFR surface densities in their central regions \citep{Kennicutt2005}. These systems lend themselves to being classified as present-day starbursts because of the observational impact made by their centrally concentrated SF, particularly when classified with short timescale star formation tracers (e.g., H$\alpha$ emission and W-R stars). In contrast, the currently more distributed starbursts (e.g., ESO 154$-$023, UGC~4483, NGC~4068) do not always meet this diverse range of observational criteria and are thus less readily identified as starbursts. The range in the spatial extent of the starbursts highlights a bias towards classifying galaxies with centralized SF as starbursts over galaxies with more distributed SF. This bias is complicated by the fact that there is no absolute threshold or criterion for classifying a galaxy as a starburst system. 

\begin{figure}
\begin{center}
\vspace{-0.5in}
\includegraphics[width=\columnwidth, clip=true]{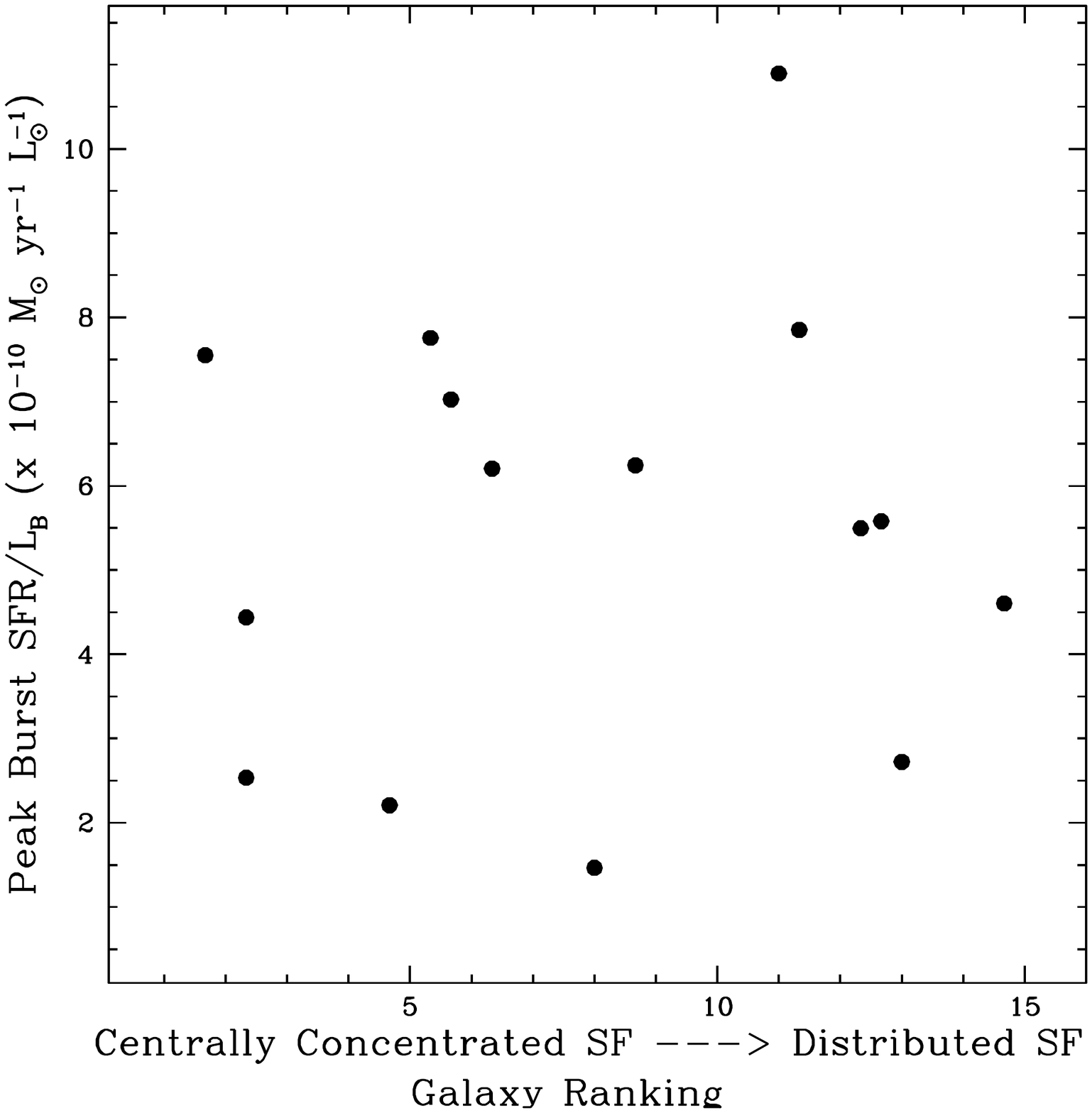}
\vspace{-0.75in}
\caption{Comparison of the galaxies ranked by their their spatial concentration of SF with the peak SFR of the starbursts, normalized by their absolute B-band luminosity in units of \lsun. The final rankings were determined by averaging the individual rankings of the three concentration metrics, from the most centrally concentrated to the most distributed. The three rankings were in general agreement, spanning a range that varied between $1-4$. The peak SFRs were taken from Paper~II and normalized to elliminate bias in the SFR due to the size of a galaxy. Uncertainties in the normalized SFRs are smaller than the individual points shown. As seen in the Figure, there is no correlation between the spatial extent of the starburst and the SFR normalized by luminosity, nor was there a correlation with the total SFR.}
\label{fig:correlate_rankings}
\end{center}
\end{figure}

Some of the most well-known starbursts in the sample that have the highest degree of centralized SF also have the highest SFRs, suggesting a possible correlation between the SFR in a starburst and the spatial extent of the SF. To explore this possible connection, we ranked the galaxies according to the three concentration metrics and then compared the average of these three rankings to the peak SFR of the starbursts. The peak SFRs (Paper~II) were normalized by the absolute B-band luminosity (in units of \lsun) to elliminate any bias in the SFR due to galaxy size. In Figure~\ref{fig:correlate_rankings}, we present this comparison with the most centralized starbursts ranked with smaller numbers and the most distibuted ranked with larger numbers. We find no correlation between the spatial extent of the SF and the normalized SFR of starbursts. 

\section{Conclusions}\label{conclusions}
We have used optical imaging of resolved stellar populations obtained from the $HST$ data archive to study the spatial distribution of SF in fifteen nearby starburst dwarf galaxies. We measure the concentration of SF over three different timescales using (1) the location of young BHeB stars from the V and I band observations, (2) newly derived SFHs for central and outer regions of the galaxies, and (3) an estimate of the fraction of stellar mass created in the central regions. The three metrics show that the spatial extent of SF lies on a continuum in these galaxies both over the most recent $100-250$ Myr, and over the life of the starburst events. SF is seen to be highly concentrated in a some of the sample (e.g., NGC 4163, NGC~5253, and NGC~6789), distributed in other systems (e.g., NGC~4068, ESO~154$-$023, NGC~784), and somewhere in between for the remaining galaxies. The range in spatial extent in the starbursts did not correlate with the peak SFR of the starbursts.

These new findings on the greater spatial extent of these starbursts, combined with the recent results measuring starbursts to last hundreds of Myr (Paper~II), provide a new perspective on the bursting mode of SF. In contrast to the often reported or assumed smaller, centralized spatial scales and short timescales (10 Myr) of starbursts, these results show that starbursts can be a global phenomenon. In addition, the range in spatial extent of the starbursts, coupled with the different observational signatures of centralized versus distributed SF, highlights a bias towards classifying galaxies with centralized SF as starbursts over galaxies with more distributed SF. This bias is complicated by the fact that there is no absolute threshold or criterion for classifying a galaxy as a starburst system.

While not a volume limited sample of starburst dwarf galaxies, these results provide context for understanding the characteristics of starbursts in low-mass galaxies. Any trigger mechanisms for the elevated levels of SF in our sample must account for these characteristics. The physical mechanism governing the spatial and temporal extent of these starbursts was not investigated, but is likely based both on the starburst trigger mechanism and the equation of state of the gas, including the degree of turbulence present in the ISM, in each system.

\section{Acknowledgments}
Support for this work was provided by NASA through grants AR-10945 and AR-11281 from the Space Telescope Science Institute, which is operated by Aura, Inc., under NASA contract NAS5-26555, and through a ROSES grant (No.~NNX10AD57G). E.~D.~S. is grateful for partial support from the University of Minnesota. The authors thank the anonymous referee for helpful comments that improved the manuscript. K.~B.~W.~M. gratefully acknowledges Matthew, Cole, and Carling for their support. This research made use of NASA's Astrophysical Data System and the NASA/IPAC Extragalactic Database (NED) which is operated by the Jet Propulsion Laboratory, California Institute of Technology, under contract with the National Aeronautics and Space Administration. 

{\it Facilities:} \facility{Hubble Space Telescope}

\appendix
\section{Maps of the Recent SF Traced by BHeB Stars}\label{appendix_figs}
In Figure~\ref{fig:galaxies}, we present the CMDs, BHeB luminosity functions, and maps of the sites of recent SF for four luminosity or age bins for the entire sample. This analysis follows the prescription described in \S\ref{heb_tracers}. In systems with higher metallicity (i.e., higher levels of extinction) and photometric crowding, BHeB stars only as old as 100 Myr can be unambiguously separated in the CMDs (e.g., NGC~625, NGC~5253, NGC~1569, NGC~4449), whereas BHeB stars up to 250 Myr in age could be separated in other systems (e.g., UGC~9128, UGC~4483). The scaling of the asterisks was uniquely determined for each system depending on the range of SFRs measured over the time periods studied; the size of the asterisks cannot be compared across systems. The preprint version of this work contains low resolution color figures; the published version contains high resolution color figures.

\begin{figure}
\begin{center}
\includegraphics[width=\columnwidth, clip=true]{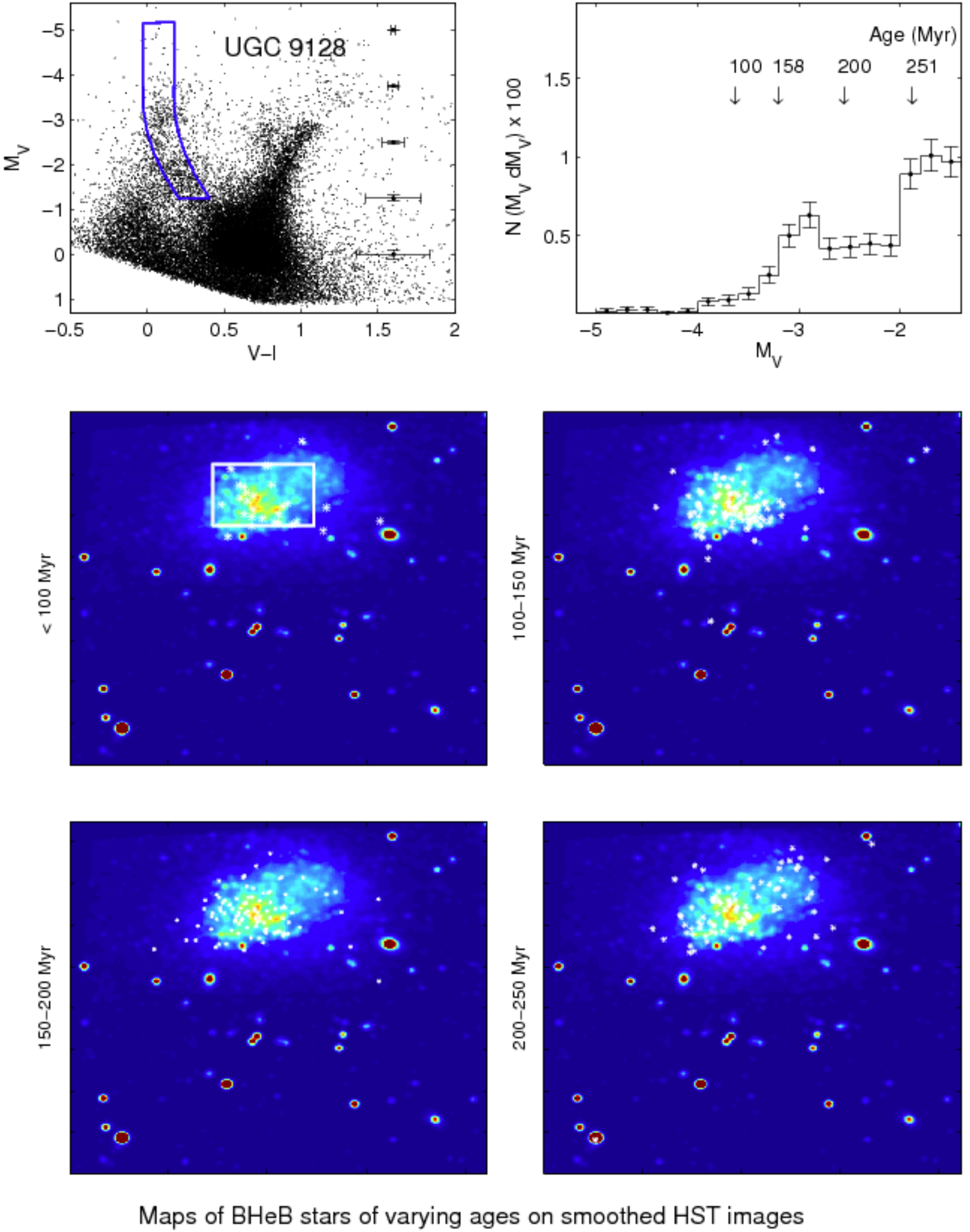}
\caption{Top left panel: CMD of the galaxy with the region containing BHeB stars used in the analysis marked in blue. Top right panel: The luminosity function of the BHeB stars identified in the CMD. Middle and lower panels: The location of the BHeB stars of four different age ranges overlaid on smoothed V band HST images of the galaxy. The area of the points is scaled by the SFR enabling a direct comparison between age bins. The white box drawn on the first image encompasses the area used in reconstructing the SFH of the central region. The region outside of this box comprises the outer region used in our analysis. Note that the HST images are not shown in WCS orientations. UGC~9128 hosts a fossil burst that ended in the last $<15$ Myr.}
\label{fig:galaxies}
\end{center}
\end{figure}

\clearpage
\begin{figure}
\begin{center}
\figurenum{\ref{fig:galaxies}}
\includegraphics[width=\columnwidth, clip=true]{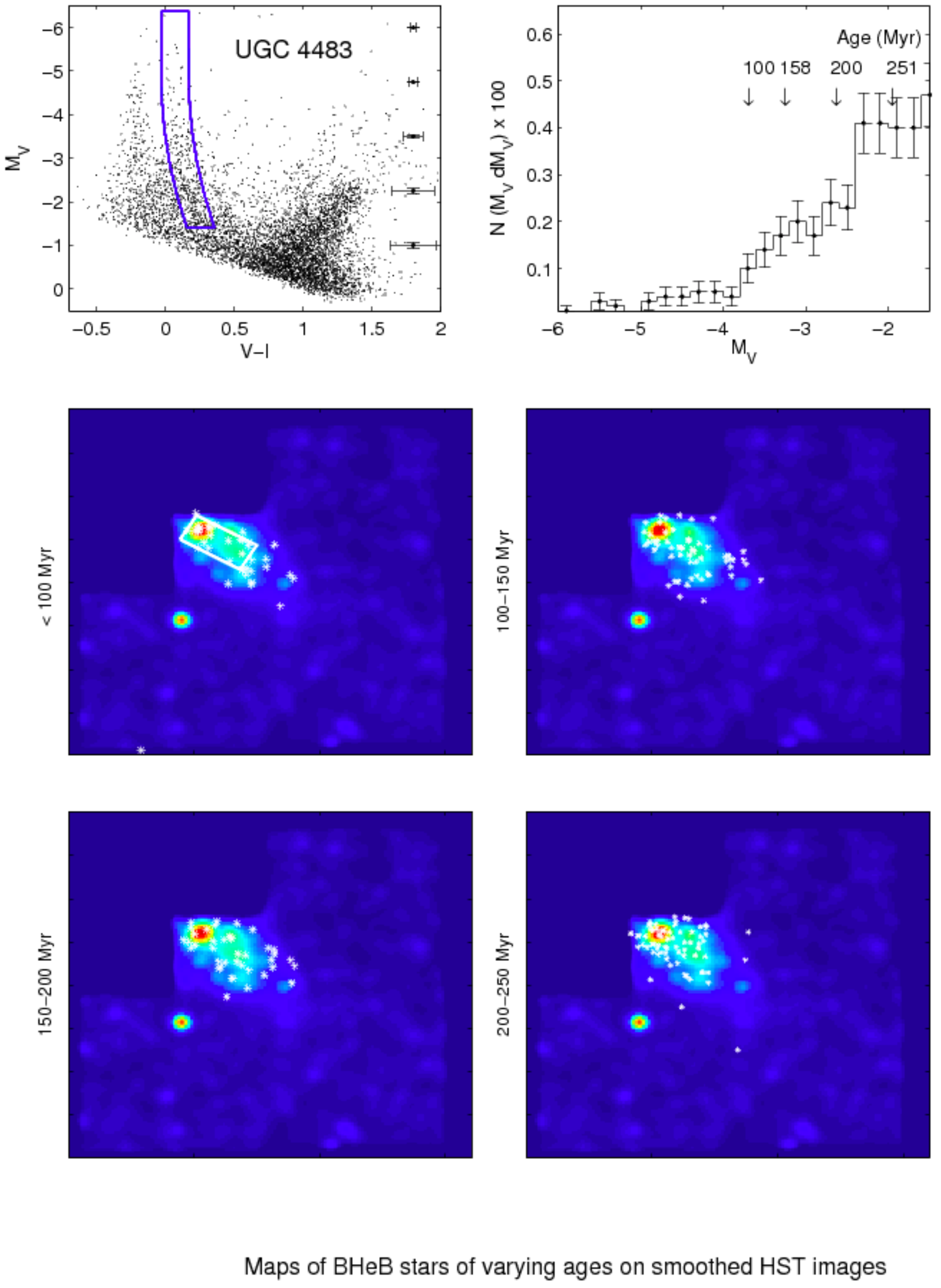}
\caption{\textit{Cont.} Note that the WFPC2 PC chip was positioned on the center of the galaxy,  thus the analysis was performed within the ``chevron'' shape of the field of view.}
\end{center}
\end{figure}

\clearpage
\begin{figure}
\begin{center}
\figurenum{\ref{fig:galaxies}}
\includegraphics[width=\columnwidth, clip=true]{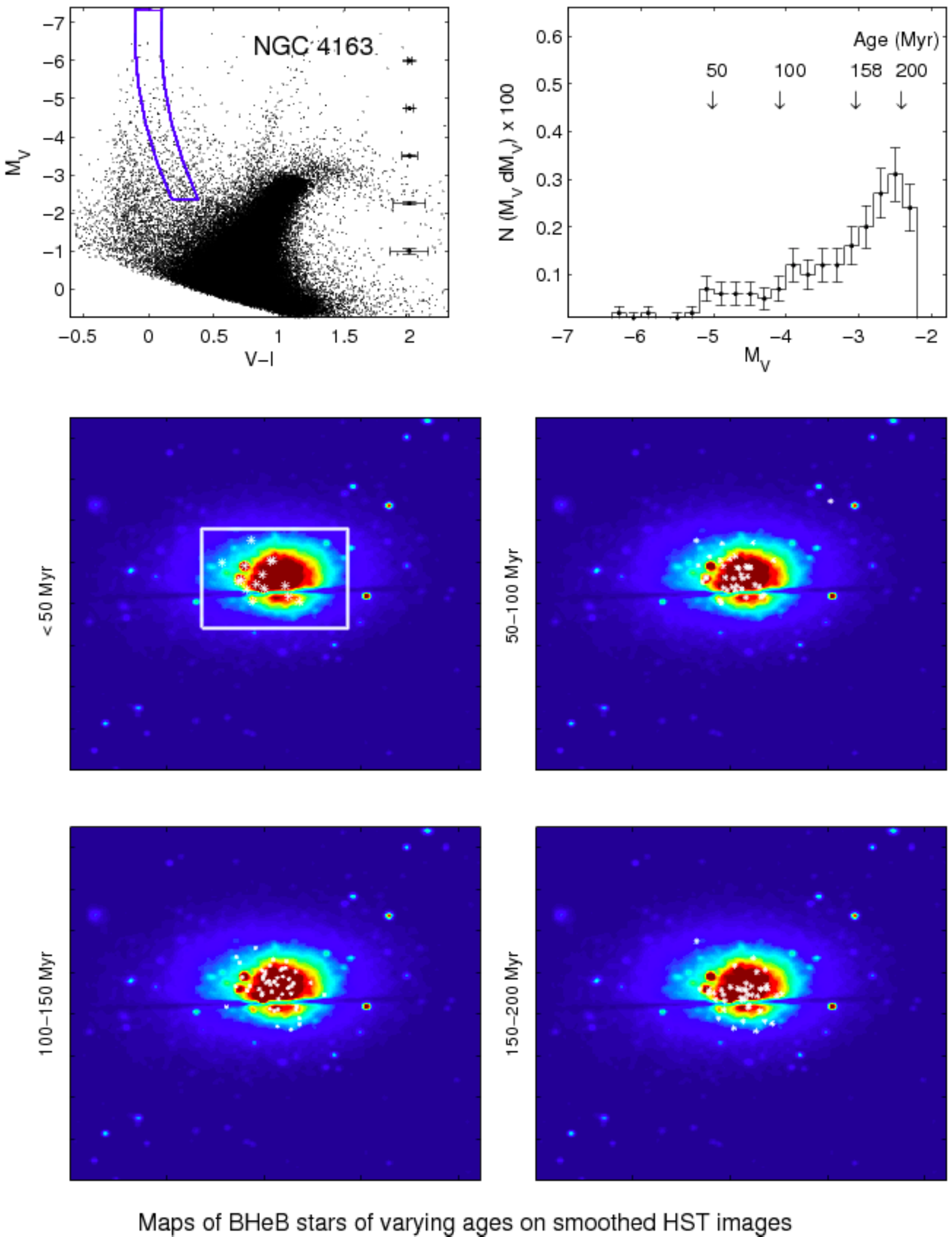}
\caption{\textit{Cont.} NGC~4163 hosts a fossil starburst that ended $\sim100$ Myr ago. Thus, the sites of SF shown in the middle panels (i.e., t$<50$ Myr and 100$<$t$<$50 Myr) are not considered part of the starburst event. We show all four time bins for completeness, but only the spatial distribution of the SF between $100-200$ Myr is considered in our analysis.}
\end{center}
\end{figure}

\clearpage
\begin{figure}
\begin{center}
\figurenum{\ref{fig:galaxies}}
\includegraphics[width=\columnwidth, clip=true]{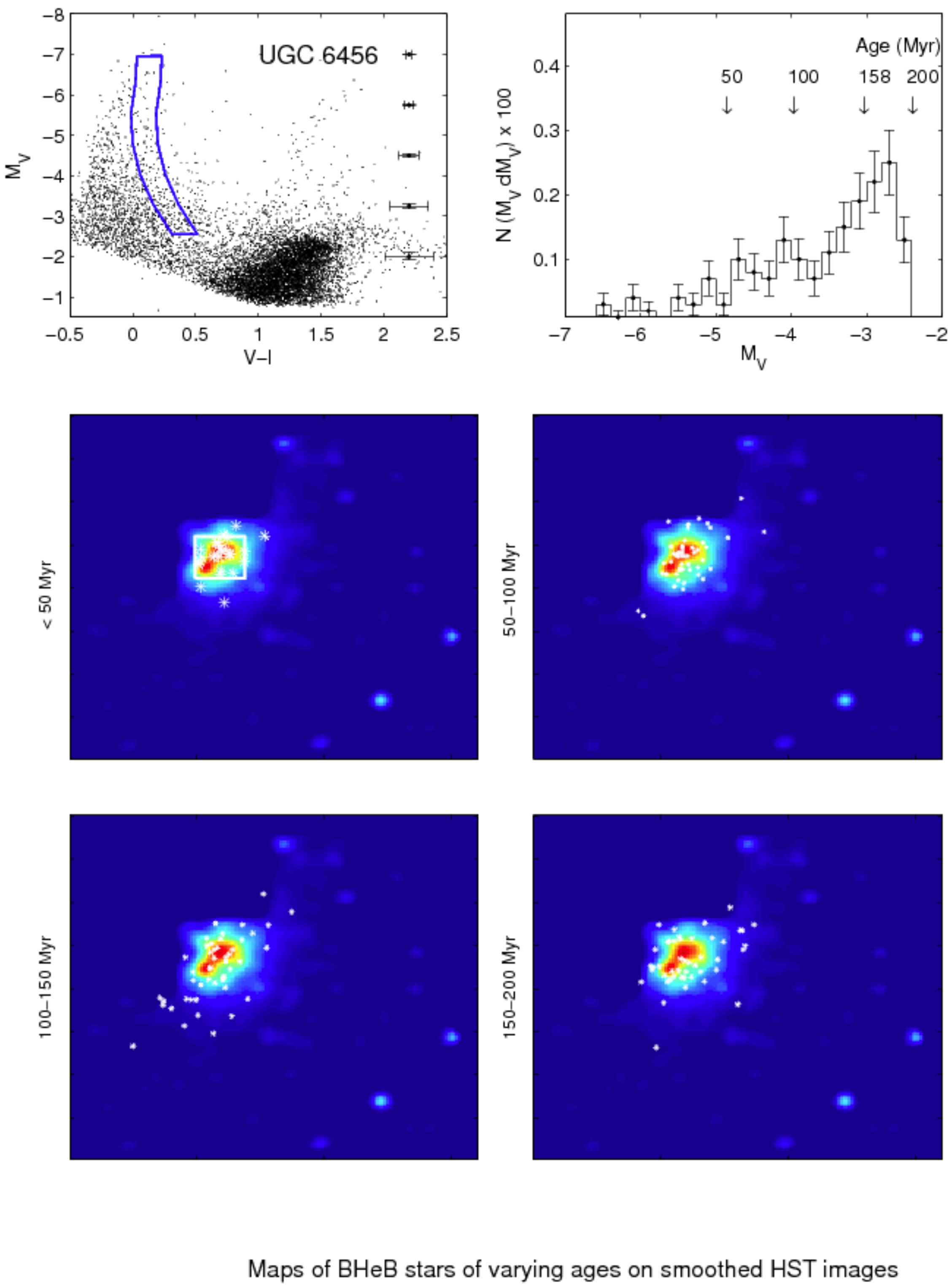}
\caption{\textit{Cont.} Note that the WFPC2 PC chip was positioned on the center of the galaxy,  thus the analysis was performed within the ``chevron'' shape of the field of view.}
\end{center}
\end{figure}

\clearpage
\begin{figure}
\begin{center}
\figurenum{\ref{fig:galaxies}}
\includegraphics[scale=0.75, clip=true]{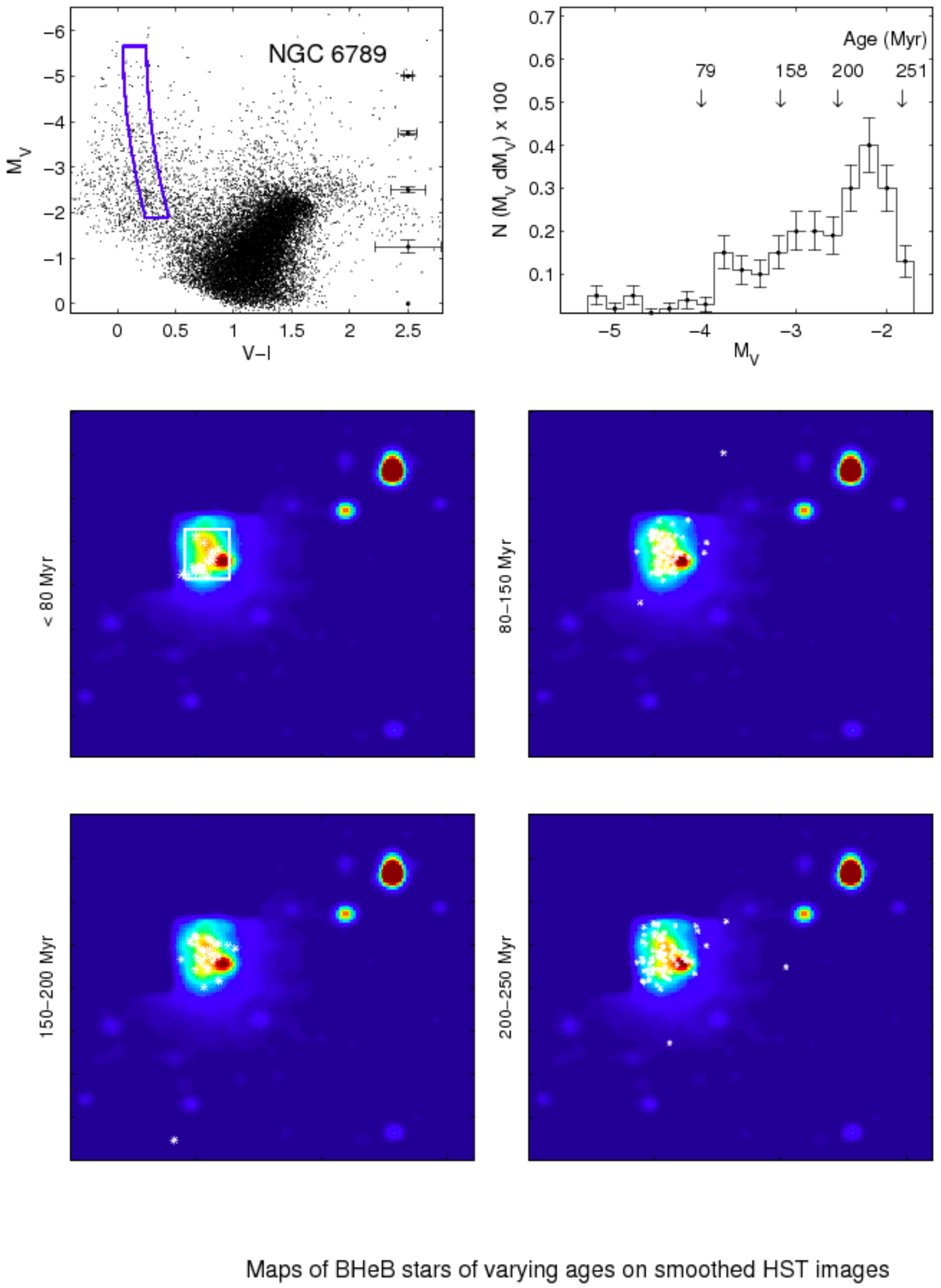}
\caption{\textit{Cont.} NGC~6789 hosts a fossil starburst that ended $\sim80$ Myr ago. Thus, the sites of SF shown in the $<80$ Myr panel are not considered part of the starburst event. We show all four time bins for completeness, but only the spatial distribution of the SF between $80-250$ Myr is considered in our analysis. Also note that the WFPC2 PC chip was positioned on the center of the galaxy, thus the analysis was performed within the ``chevron'' shape of the field of view.}
\end{center}
\end{figure}

\clearpage
\begin{figure}
\begin{center}
\figurenum{\ref{fig:galaxies}}
\includegraphics[width=\columnwidth, clip=true]{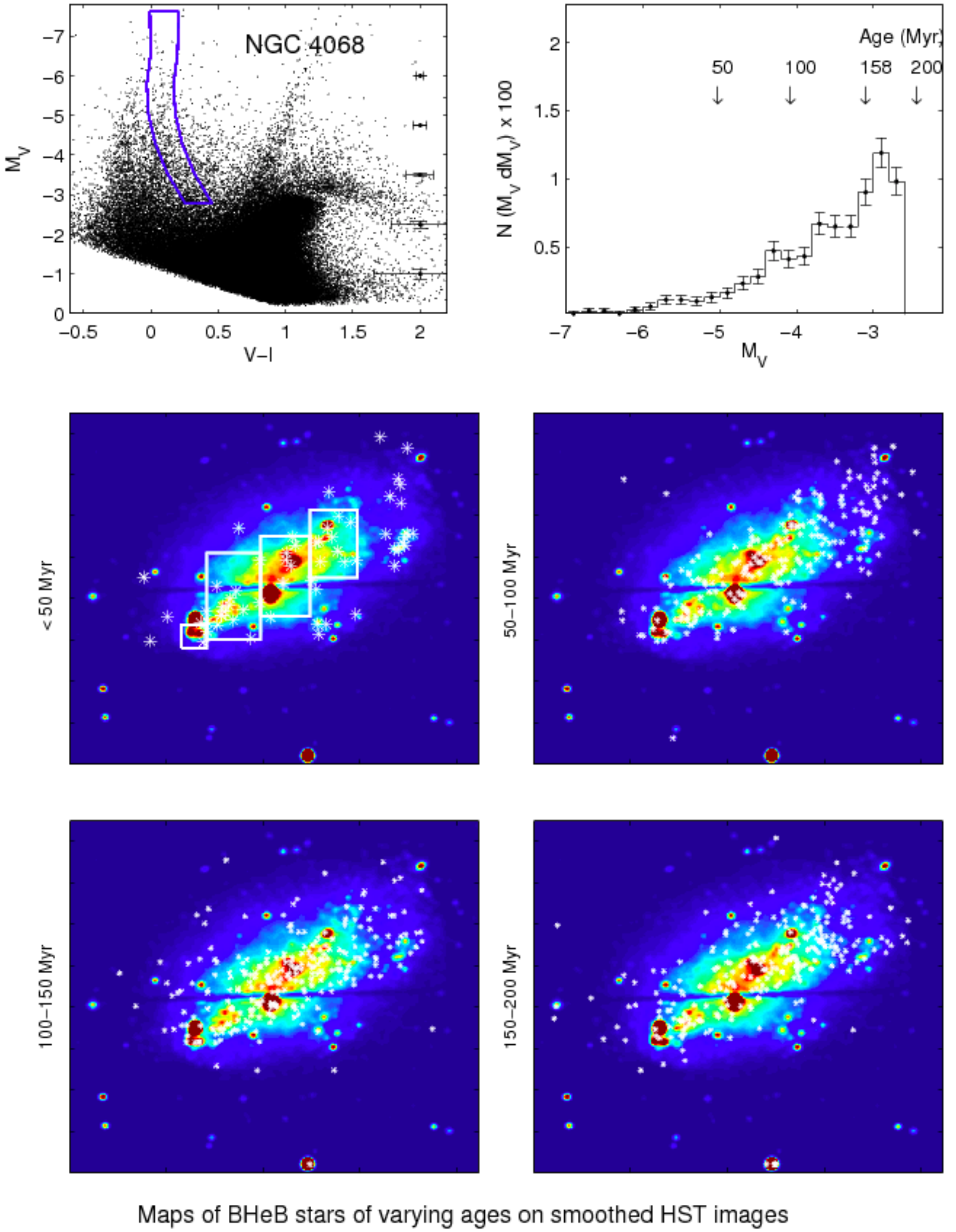}
\caption{\textit{Cont.}}
\end{center}
\end{figure}

\clearpage
\begin{figure}
\begin{center}
\figurenum{\ref{fig:galaxies}}
\includegraphics[width=\columnwidth, clip=true]{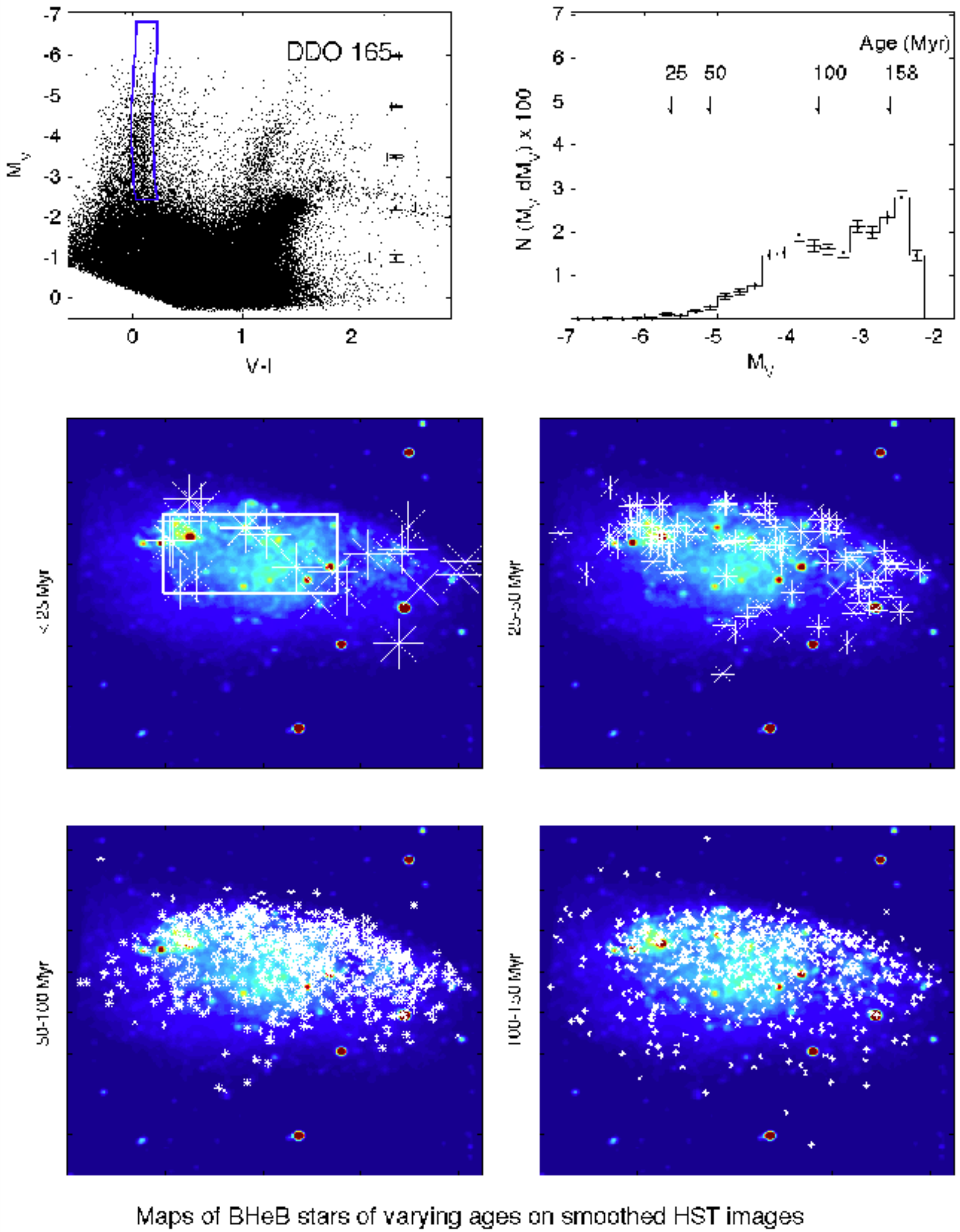}
\caption{\textit{Cont.} Note the large asterisks reflect a sustained, high level of SF over the time bins studied.}
\end{center}
\end{figure}

\clearpage
\begin{figure}
\begin{center}
\figurenum{\ref{fig:galaxies}}
\includegraphics[width=\columnwidth, clip=true]{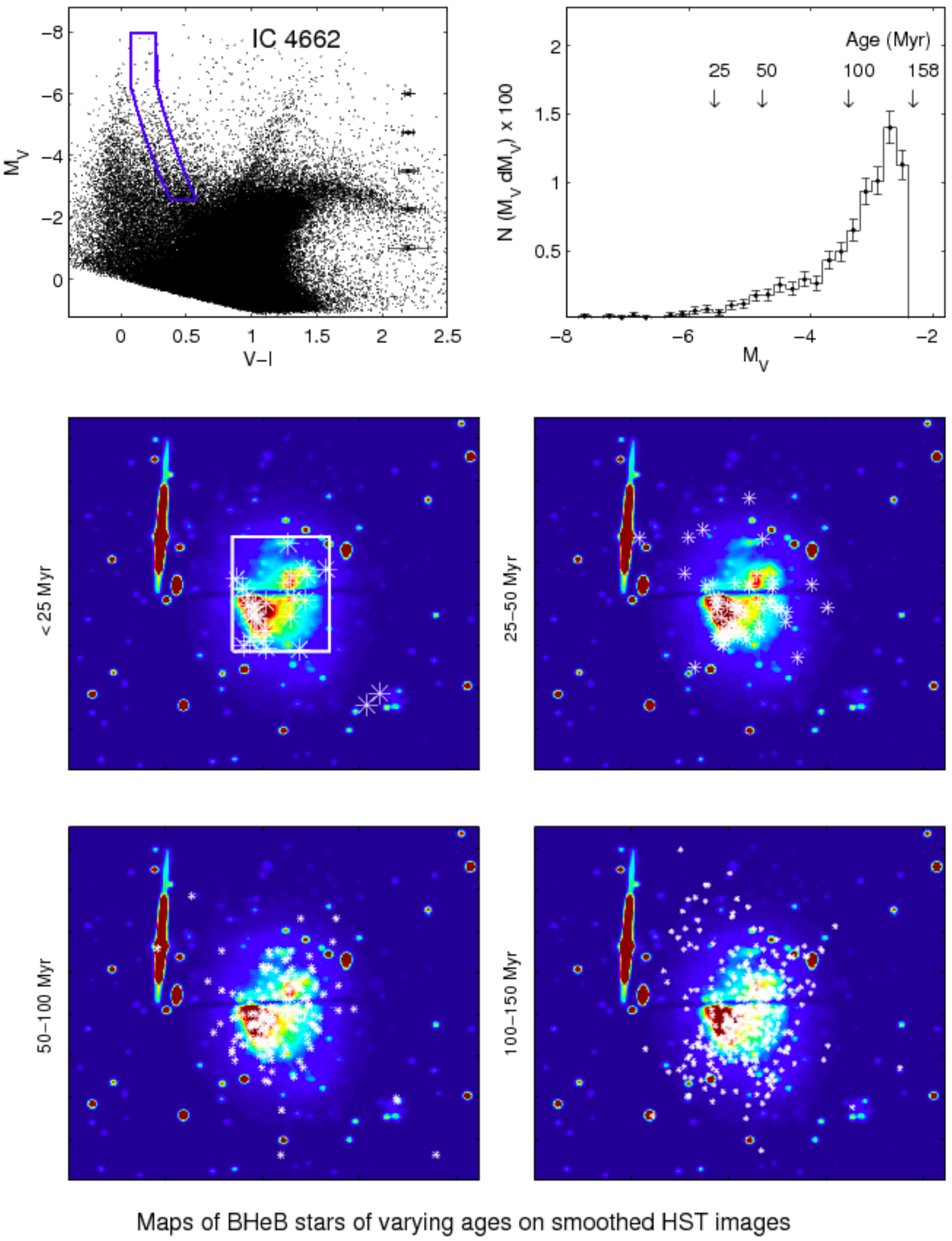}
\caption{\textit{Cont.}}
\end{center}
\end{figure}

\clearpage
\begin{figure}
\begin{center}
\figurenum{\ref{fig:galaxies}}
\includegraphics[width=\columnwidth, clip=true]{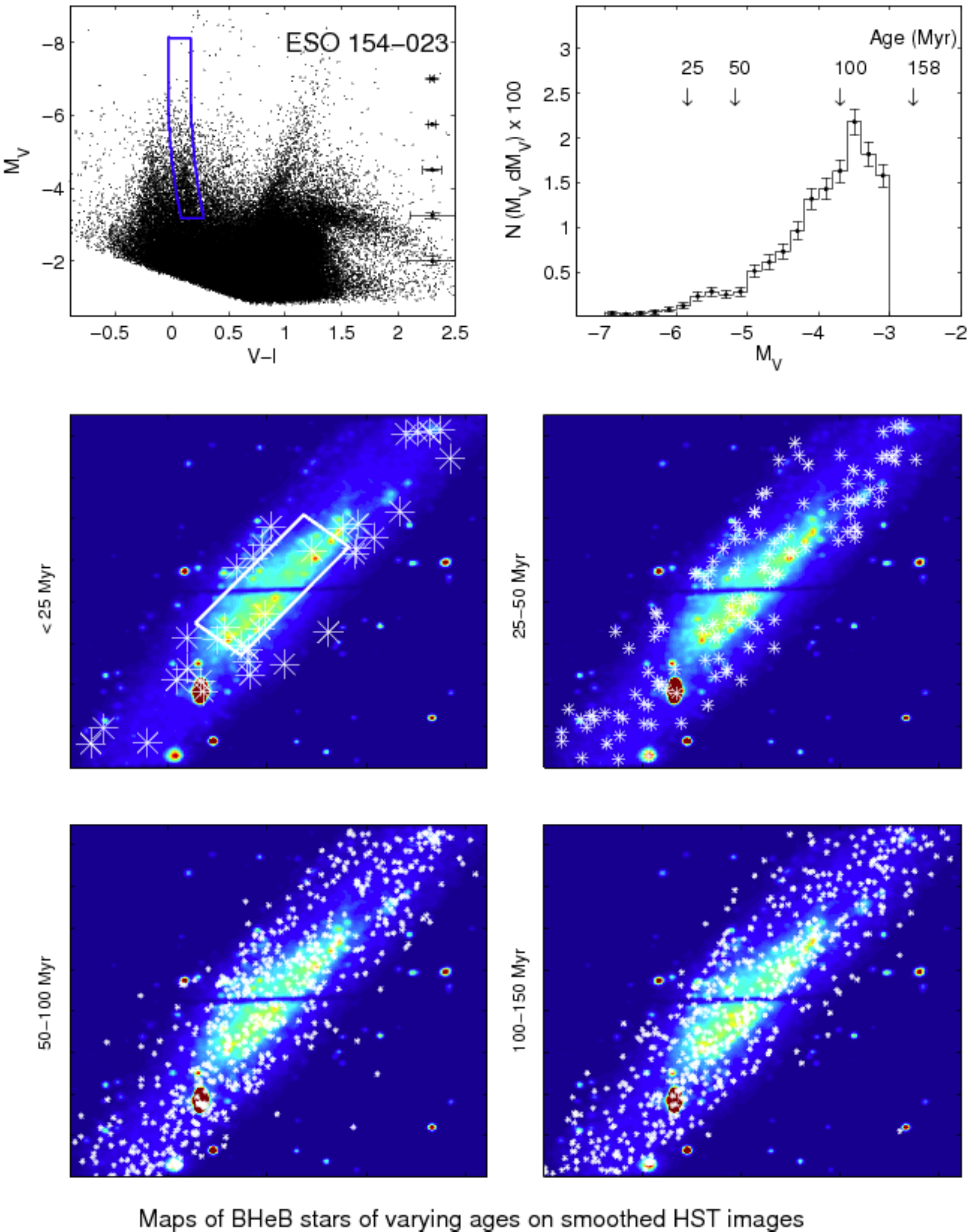}
\caption{\textit{Cont.} Note the large asterisks reflect a sustained, high level of SF over the time bins studied.}
\end{center}
\end{figure}

\clearpage
\begin{figure}
\begin{center}
\figurenum{\ref{fig:galaxies}}
\includegraphics[width=\columnwidth, clip=true]{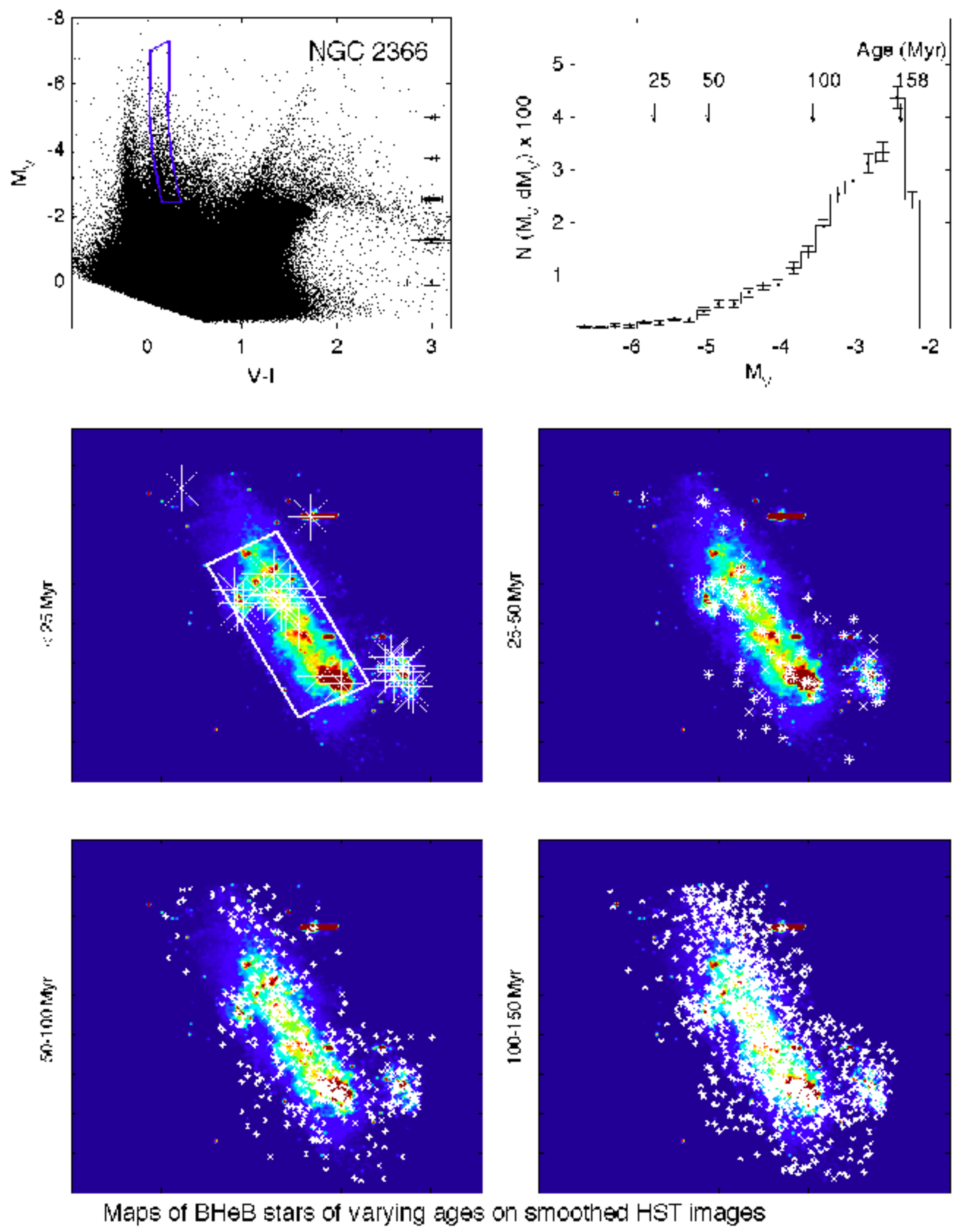}
\caption{\textit{Cont.} Note the large asterisks reflect a sustained, high level of SF over the time bins studied.}
\end{center}
\end{figure}

\clearpage
\begin{figure}
\begin{center}
\figurenum{\ref{fig:galaxies}}
\includegraphics[scale=0.75, clip=true]{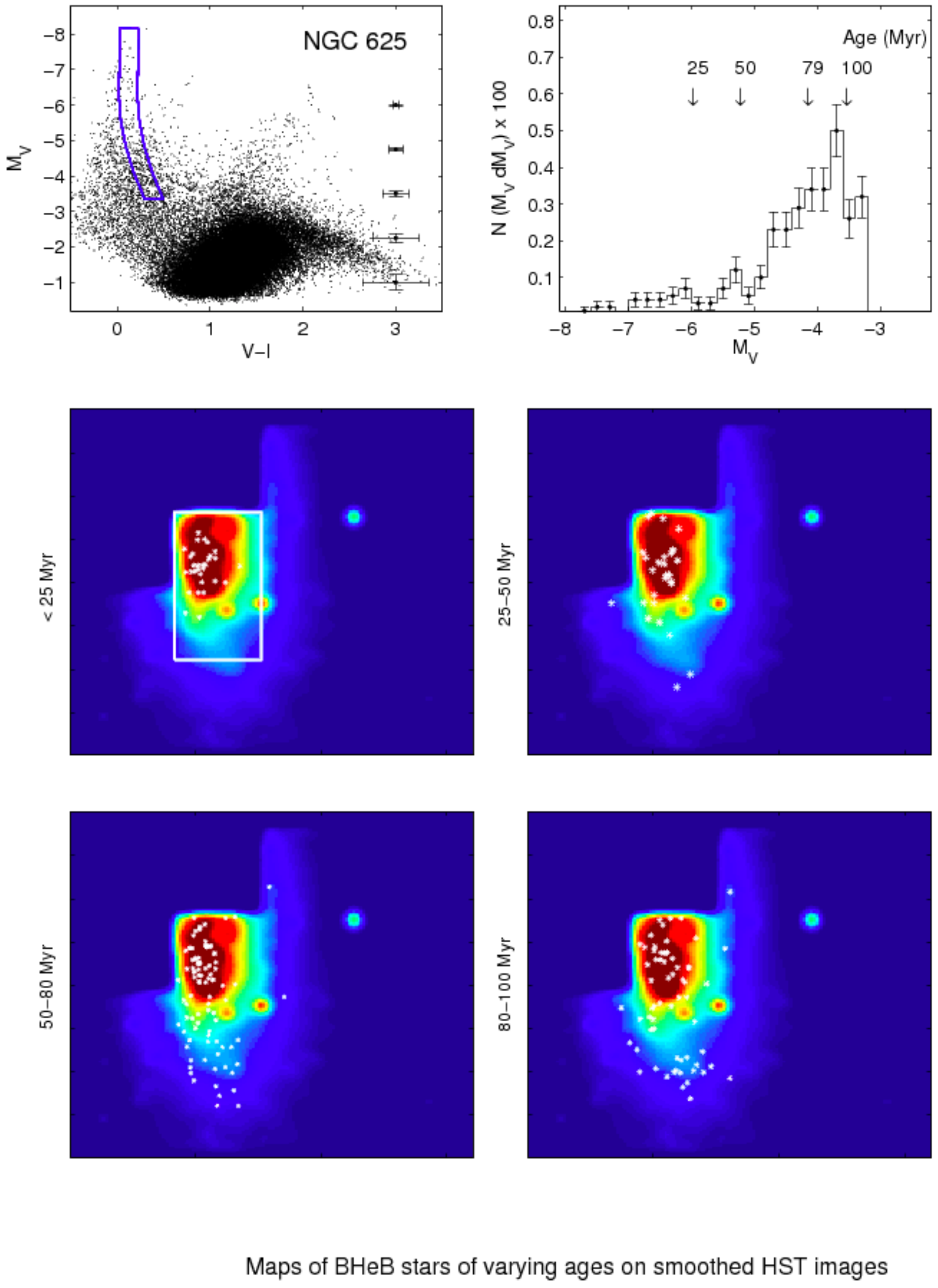}
\caption{\textit{Cont.} Note that while NGC~625 hosts a fossil burst than ended $<35$ Myr ago when averaged over the entire field of view, the central regions in this system show ongoing levels of SF. Thus, we include the SF over the most recent 35 Myr for completeness in the analysis of this starburst event.}
\end{center}
\end{figure}

\clearpage
\begin{figure}
\begin{center}
\figurenum{\ref{fig:galaxies}}
\includegraphics[width=\columnwidth, clip=true]{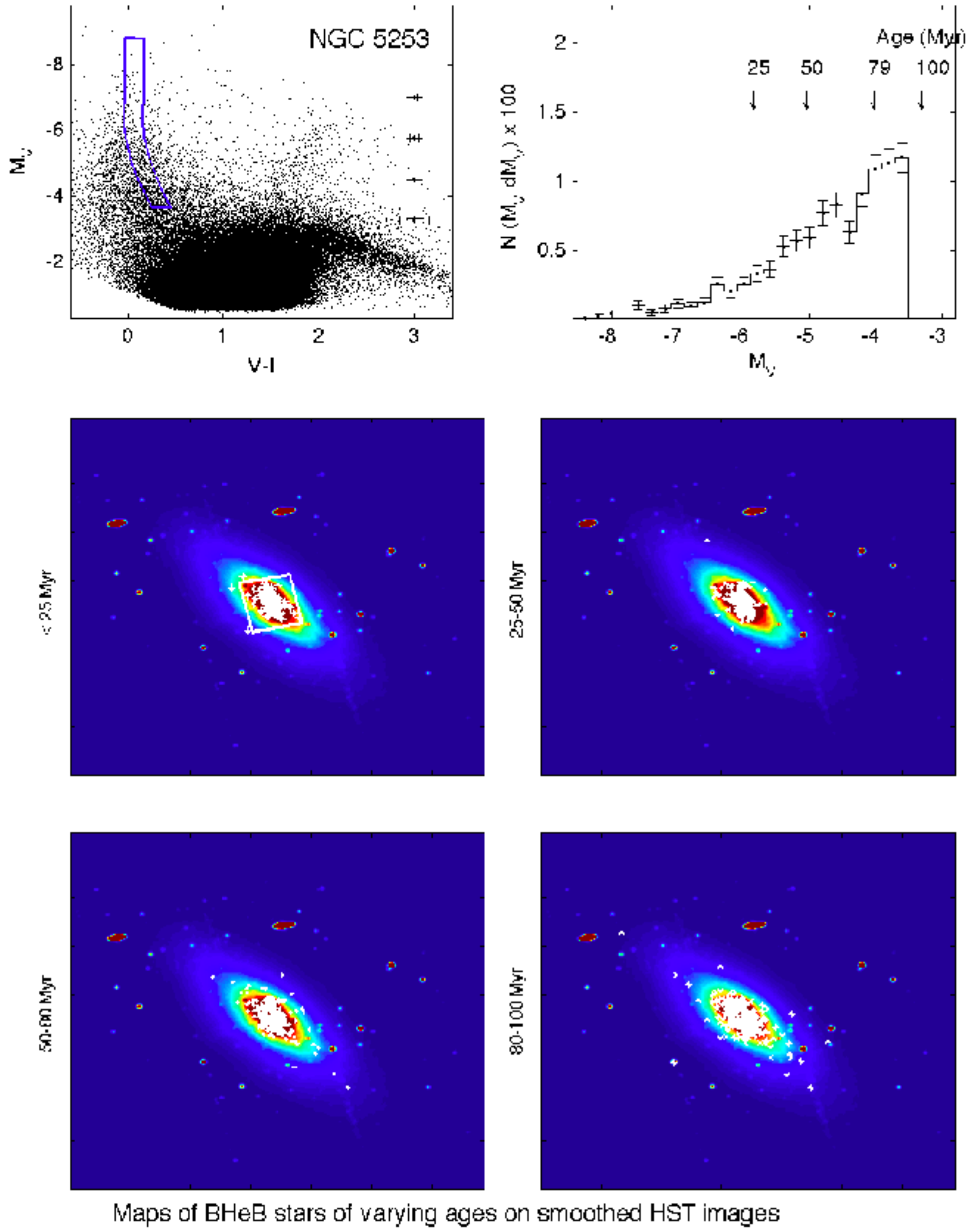}
\caption{\textit{Cont.}}
\end{center}
\end{figure}

\clearpage
\begin{figure}
\begin{center}
\figurenum{\ref{fig:galaxies}}
\includegraphics[width=\columnwidth, clip=true]{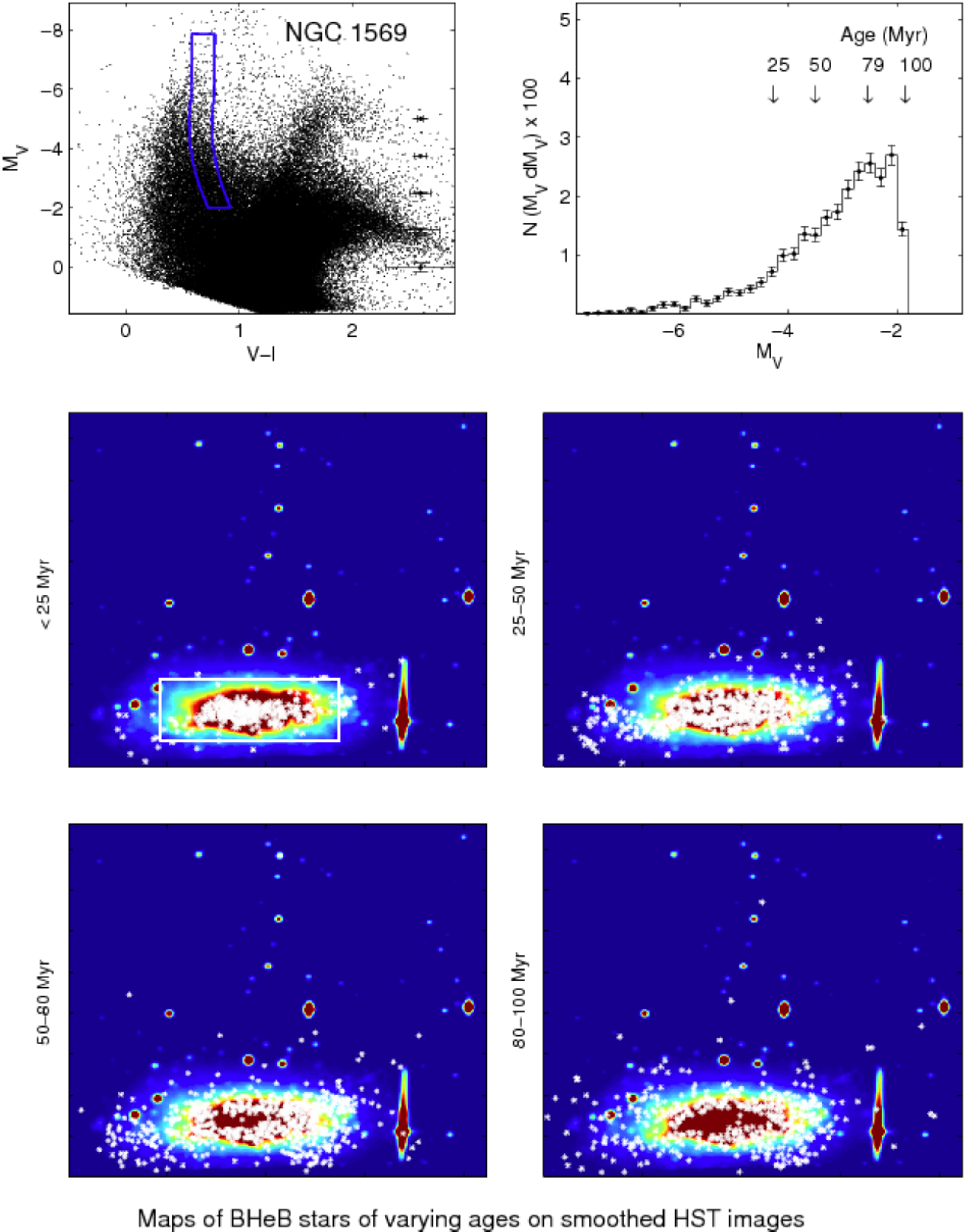}
\caption{\textit{Cont.}}
\end{center}
\end{figure}

\clearpage
\begin{figure}
\begin{center}
\figurenum{\ref{fig:galaxies}}
\includegraphics[width=\columnwidth, clip=true]{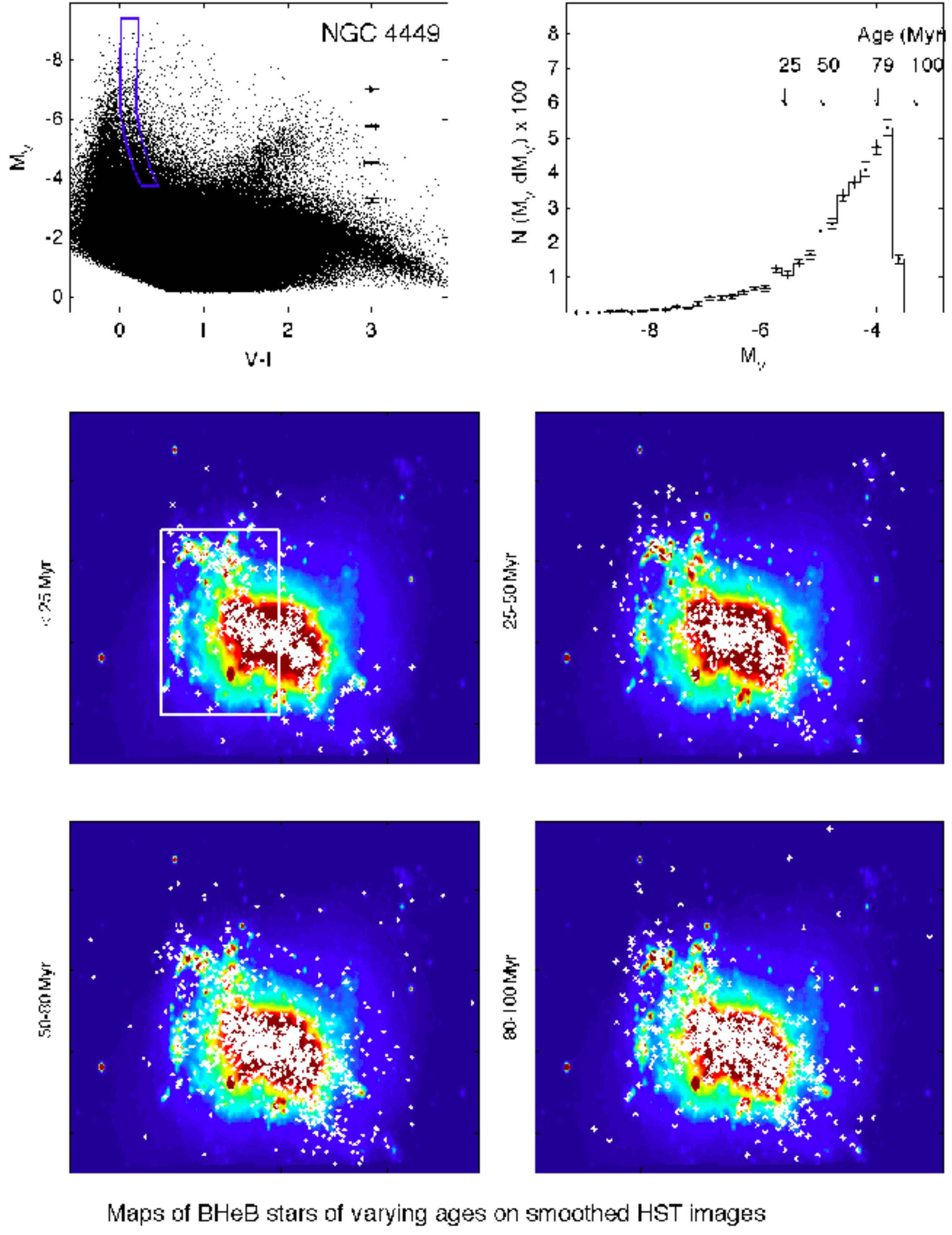}
\caption{\textit{Cont.}}
\end{center}
\end{figure}

\clearpage
%%tab:properties
\input{tab1}

%%tab:spatial_extent
\input{tab2}

\end{document}

%% file: tab1.tex
%%
%% McQuinn et al., Spatial Characteristics of  Starbursts
%% table 1

\begin{deluxetable}{lcrrrr}
\tabletypesize{\scriptsize}
\tablewidth{0pt}
\tablecaption{The Galaxy Sample and BHeB Star Timescales\label{tab:properties}}
\tablecolumns{6}
\tablehead{
\colhead{}				&
\colhead{M$_B$}				&
\colhead{D$_{25}$ Area}			&
\colhead{Deproj. D$_{25}$ Area}		&
\colhead{Duration}			&
\colhead{Max. Meas. Age of}		\\
\colhead{Galaxy}			&
\colhead{(mag)}				&
\colhead{in FOV (\%)}			&
\colhead{in FOV (kpc$^2$)}		&
\colhead{of Burst (Myr)}		&
\colhead{BHeB stars (Myr)}		\\
\colhead{(1)}				&	
\colhead{(2)}				&	
\colhead{(3)}				&	
\colhead{(4)}				&	
\colhead{(5)}				&	
\colhead{(6)}					
}

\startdata
UGC 9128	& -12.45 & 100\% & 1.2   & 1300$\pm$300		& 250 \\
UGC 4483	& -12.68 & 100\% & 1.0   & $>$810$\pm$190	& 250 \\
NGC 4163	& -13.75 & 100\% & 3.2   & 460$\pm$70		& 200 \\
UGC 6456 	& -13.85 & 100\% & 3.0   & $>$570$\pm$60	& 200 \\
NGC 6789 	& -14.60 & 100\% & 1.7   & 480$\pm$70		& 250 \\
NGC 4068 	& -14.96 & 100\% & 19.5  & $>$459$\pm$50	& 200 \\
DDO 165 	& -15.19 & 100\% & 18.   & $>$1300$\pm$300	& 158 \\
IC 4662 	& -15.39 & 100\% & 2.2   & $>$450$\pm$50	& 158 \\
ESO 154-023 	& -16.21 & 63\%  & 120   & $>$450$\pm$50	& 158 \\
NGC 2366	& -16.33 & 96\%  & 40.   & $>$450$\pm$50	& 158 \\
NGC 625 	& -16.26 & 63\%  & 35.   & 450$\pm$50		& 100 \\
NGC 784 	& -16.78 & 73\%  & 64.   & $>$450$\pm$50	& 158 \\
NGC 5253 	& -16.98 & 100\% & 28.   & $>$450$\pm$50	& 100 \\
NGC 1569 	& -17.76 & 87\%  & 3.0   & $>$450$\pm$50	& 100 \\
NGC 4449 	& -18.02 & 87\%  & 40.   & $>$450$\pm$50	& 100 \\

\enddata
\tablecomments{Column (1) Galaxy name. Column (2) B$-$band luminosity of the galaxy. Column (3) Indicates the percentage of the D$_{25}$ area \citep{Karachentsev2004} in the observational field of view (F.O.V.). Column (4) Measures the deprojected area in the F.O.V. out to the D$_{25}$ area. Column (5) The starburst durations reported in Paper~II. Lower limits indicate that starbursts are on-going in these systems. Total starburst durations are measured in 4 fossil burst systems. Column (6) The timescale over which it is possible to unambiguously select the BHeB from the optical CMD with a maximum age of 250 Myr. The maximum age of 250 Myr was chosen as this timescale approximates the lower limit for which stellar structure has been shown to exist in dwarf galaxies \citep{Bastian2011}. Diffusion of stars from their birth sites becomes more important at longer timescales, thus reducing the accuracy of tracing sites of recent SF with the distribution of HeB stars older than $\sim250$ Myr.}

\end{deluxetable}

%% file: tab2.tex
%%
%% McQuinn et al., Spatial Characteristics of  Starbursts
%% table 2

\begin{deluxetable}{lccrcr}
\tabletypesize{\scriptsize}
%\rotate
\tablewidth{0pt}
\tablecaption{The Spatial Extent of the Bursts\label{tab:spatial_extent}}
\tablecolumns{6}
\tablehead{
\colhead{}						&
\multicolumn{3}{c}{Most Recent $\sim100-250$ Myr}	&
\multicolumn{2}{c}{Lifetime of Burst}			\\
\colhead{}						&
\colhead{Concentration Ratio of}			&
\colhead{Ratio of Central to Outer}			&
\colhead{Fraction M$_{*}$ in}				&
\colhead{Ratio of Central to Outer}			&
\colhead{Fraction M$_{*}$ in}				\\
\colhead{}						&
\colhead{BHeB/RGB stars (\%)}				&
\colhead{Birthrate Parameters}				&
\colhead{Central Region}				&
\colhead{Birthrate Parameters}				&
\colhead{Central Region}				\\
\hline \hline						\\
\colhead{Galaxy}					&
\colhead{(2)}						&
\colhead{(3)}						&
\colhead{(4)}						&
\colhead{(5)}						&
\colhead{(6)}						
}

\startdata
UGC 9128	& 44   &  4.3  $\pm$  0.4    &  69 $\pm$  7	 & 1.3 $\pm$  0.1    &   40 $\pm$  4	 \\
UGC 4483	& 71   &  4.0  $\pm$  0.4    &  46 $\pm$  4	 & 1.6 $\pm$  0.1    &   25 $\pm$  5	\\
NGC 4163	& 21   &  0.0  $\pm$  0.0    &  96 $\pm$  10	 & 1.3 $\pm$  0.1    &   70 $\pm$  13	\\
UGC 6456 	& 20   &  16.0 $\pm$  2.3    &  58 $\pm$  3	 & 4.8 $\pm$  0.5    &   29 $\pm$  6	  \\
NGC 6789 	& ...  &  16.5 $\pm$  0.1    &  85 $\pm$  3	 & 4.3 $\pm$  0.1    &   56 $\pm$  13	\\
NGC 4068 	& 97   &  2.1  $\pm$  0.1    &  51 $\pm$  2	 & 0.1 $\pm$  0.1    &   37 $\pm$  1	  \\
DDO~165 	& 65   &  2.5  $\pm$  0.1    &  46 $\pm$  2	 & 1.3 $\pm$  0.1    &   32 $\pm$  2	\\
IC~4662 	& 25   &  5.8  $\pm$  0.1    &  78 $\pm$  3	 & 2.7 $\pm$  0.1    &   59 $\pm$  8	 \\
ESO 154-023 	& 100  &  0.9  $\pm$  0.1    &  27 $\pm$  2	 & 1.1 $\pm$  0.1    &   30 $\pm$  4	\\
NGC 2366	& 85   &  1.5  $\pm$  0.2    &  61 $\pm$  6	 & 1.0 $\pm$  0.1    &   52 $\pm$  4	    \\
NGC 625 	& 10   &  4.8  $\pm$  0.1    &  63 $\pm$  5	 & 3.0 $\pm$  0.1    &   51 $\pm$  9	\\
NGC 784 	& 54   &  1.3  $\pm$  0.1    &  25 $\pm$  1	 & 0.8 $\pm$  0.1    &   19 $\pm$  2	  \\
NGC 5253 	& 13   &  0.0  $\pm$  0.0    &  94 $\pm$  6	 & 4.0 $\pm$  0.1    &   59 $\pm$  6	 \\
NGC 1569 	& 15   &  12.1 $\pm$  0.1    &  83 $\pm$  4	 & 4.7 $\pm$  0.1    &   67 $\pm$  8	\\
NGC 4449 	& 28   &  7.6  $\pm$  0.1    &  84 $\pm$  4	 & 3.4 $\pm$  0.1    &   69 $\pm$  13	\\

\enddata
\tablecomments{Column (1) Galaxy name. Columns (2)$-$(4) Results are based on the SF from 0 Myr to $100-250$ Myr ago (see Table~\ref{tab:properties}, Column 6) except for NGC~4163 and NGC~6789. In these two post-burst galaxies, the SF was measured from $100-200$ Myr and $80-250$ Myr ago respectively, corresponding to time periods of actively bursting SF. Column (2) Comparison is based on the number density of the BHeB to the RGB stars. Uncertainties are estimated to be of order 10\%. Columns (5)$-$(6) Results are based on the SF that has occured over the duration of the starburst in each galaxy.}
\end{deluxetable}